\documentclass[11pt]{article}
\usepackage{latexsym, graphicx, amsmath, amssymb, amsthm, enumerate}
\usepackage[margin=1in]{geometry}
\usepackage[affil-it]{authblk}

\usepackage[utf8]{inputenc}
\usepackage[T1]{fontenc}
\usepackage{bm}
\usepackage{amsmath,amsfonts,amssymb,multirow}
\usepackage{multicol}

\usepackage{epsfig}
\usepackage{graphics, sidecap, wrapfig}
\usepackage{float}

\usepackage[dvipsnames,usenames]{color}
\usepackage[colorlinks=true,urlcolor=Blue,citecolor=Green,linkcolor=BrickRed]{hyperref}
\usepackage{cleveref}
\usepackage[usenames,dvipsnames]{xcolor}
\usepackage{thm-restate}
\usepackage[depth=subsection]{bookmark}
\usepackage{hhline}
\usepackage{appendix}
\usepackage{ifthen}

\def \fullver {}
\def \isarticle {}

\makeatletter

\newcommand{\lglg}[1]{\log\log{#1}}

\newcommand{\sep}[1]{$Sep_{#1}$}
\newcommand{\eps}{\varepsilon}
\newcommand{\epsm}{\varepsilon^{-1}}
\newcommand{\epsmm}{\varepsilon^{-2}}

\newcommand{\terminate}{ 
	\bibliography{dynamic}
	\end{document}
} 
 
\newcommand{\defproof}[2]{
	\@namedef{PF{#1}}{#2}

	\ifdefined \fullver
		\@nameuse{PF{#1}}
	\fi
}

\newcommand{\showproof}[1]{
	\ifdefined \fullver {}
	\else 
		\@nameuse{#1}*
		\@nameuse{PF{#1}} 
	\fi
}

	\let \mysection \section
	\let \mysubsection \subsection
	\let \mysubsubsection \subsubsection
	\ifdefined \isarticle 
		\newcommand{\qendproof}{\end{proof}}
	\else 
		\newcommand{\qendproof}{\qed\end{proof}}
	\fi
	\let \institute \affil
	\newtheorem{definition}{Definition}
	\newtheorem{lemma}{Lemma}
	\newtheorem{theorem}{Theorem}

\begin{document}

\title{Efficient Dynamic Approximate Distance Oracles for Vertex-Labeled Planar Graphs\thanks{This research was supported by the ISRAEL SCIENCE FOUNDATION (grant No. 794/13).} \thanks{For a full version of this paper, see \url{https://arxiv.org/abs/1707.02414}.}}
	\author{Itay Laish and Shay Mozes}
	\institute{Efi Arazi School of Computer Science\\
		The Interdisciplinary Center Herzliya \\
	}

	\maketitle
	\textbf{Abstract} Let $G$ be a graph where each vertex is associated with a label. A \emph{Vertex-Labeled Approximate 		Distance Oracle} is a data structure that, given  a vertex $v$ and a label $\lambda$, returns a $(1+\eps)$-approximation of the distance from $v$ to the closest vertex with label $\lambda$ in $G$. Such an oracle is {\em dynamic} if it also supports label changes.
In this paper we present three different dynamic approximate vertex-labeled distance oracles for planar graphs, all with polylogarithmic query and update times, and nearly linear space requirements.  No such oracles were previously known.

\mysection{Introduction}
Consider the following scenario.
A 911 dispatcher receives a call about a fire and needs to dispatch the closest fire truck. There are two difficulties with locating the appropriate vehicle to dispatch. First, the vehicles are on a constant move. Second, there are different types of emergency vehicles, whereas the dispatcher specifically needs a fire truck. Locating the closest unit of certain type under these assumptions is the \emph{dynamic vertex-labeled distance query problem} on the road network graph. Each vertex in this graph can be annotated with a label that represents the type of the emergency vehicle currently located at that vertex. 
	An alternative scenario where this problem is relevant is when one wishes to find a service provider (e.g., gas station, coffee shop), but different locations are open at different times of the day.

A data structure that answers distance queries between a vertex and a label, and supports label updates is called a \emph{dynamic vertex-labeled distance oracle}.
We model the road map as a planar graph, and extend previous results for the static case (where labels are fixed). We present oracles with polylogarithmic update and query times (in the number of vertices) that require nearly linear space. 

We focus on approximate vertex-labeled distance oracles for fixed parameter $\eps\geq 0$. When queried, such oracle returns at least the true distance, but not more than $(1+\eps)$ times the true distance. These are also known as stretch-$(1+\eps)$ distance oracles.
Note that, in our context, the graph is fixed, and only the vertex labels change.

\mysubsection{Related Work}\label{sec::related_work}

A seminal result on approximate vertex-to-vertex distance oracles for planar graphs is that of Thorup~\cite{Thorup04}. He presented a stretch-$(1+\eps)$ distance oracle for directed planar graphs. For any $0<\eps<1$, his oracle can be stored using $O(\eps^{-1}n\log{n}\log(nN))$ space and answer queries in $O(\lglg(nN)+\eps^{-1})$ time.
Here $N$ denotes the ratio of the largest to smallest arc length. For undirected planar graphs Thorup presented a $O(\eps^{-1}n\log{n})$ space oracle that answers queries in $O(\eps^{-1})$ time. Klein~\cite{Klein02,MSSP} independently described a stretch-$(1+\eps)$ distance
oracle for undirected graphs with the same bounds, but faster preprocessing time.

The first result for the static vertex-labeled problem for undirected planar graph is due to Li, Ma and Ning~\cite{LMN13}. They described a stretch-$(1+\eps)$ distance oracle that is based on Klein's results~\cite{Klein02}. Their oracle requires $O(\eps^{-1}n\log{n})$ space, and answers queries in $O(\eps^{-1}\log{n}\log\Delta)$ time. Here $\Delta$ is the hop-diameter of the graph, which can be $\Theta(n)$.  Mozes and Skop~\cite{MozesS15}, building on Thorup's oracle, described a stretch-$(1+\eps)$ distance oracle for directed planar graphs that can be stored using $O(\eps^{-1}n\log{n}\log(nN))$ space, and has $O(\lglg{n}\lglg{nN} +\eps^{-1})$ query time. 

Li Ma and Ning~\cite{LMN13} considered the dynamic case, but their update time is $\Theta(n\log{n})$ in the worst case. Łącki et al.~\cite{Lacki} presented a different dynamic vertex-to-label oracle for undirected planar graphs, in the context of computing Steiner trees. Their orcale requires
 $O(\sqrt{n}\log^2{n}\log D\eps^{-1})$ amortized time per update or query (in expectation), where $D$ is the stretch of the metric of the graph (could be $nN$). Their oracle however does not support changing the label of a specific vertex. It supports merging two labels, and splitting of labels in a restricted way.
To the best of our knowledge, ours are the first approximate dynamic vertex-labeled distance oracles with polylogarithmic query and update times, and the first that support directed planar graphs.

\mysubsection{Our results and techniques}
We present three approximate vertex-labeled distance oracles with polylogarithmic query and update times and nearly linear space and preprocessing times.
Our update and construction times are expected amortized due to the use of dynamic hashing.\footnote{We assume that a single comparison or addition of two numbers takes constant time.} Our solutions differ in the tradeoff between query and update times. One solution works for directed planar graphs, whereas the other two only work for undirected planar graphs.

We obtain our results by building on and combining existing techniques for the static case. All of our oracles rely on recursively decomposing the graph using shortest paths separators. Our first oracle for undirected graphs (Section~\ref{sec::fast_query}) uses uniformly spaced connections, and efficiently handles them using fast predecessor data structures. The upshot of this approach is that there are relatively few connections. The caveat is that this approach only works when working with bounded distances,
so a scaling technique~\cite{Thorup04} is required.

Our second oracle for undirected graphs (Section~\ref{sec::fast_update}) uses the approach taken by Li Ma and Ning~\cite{LMN13} in the static case. Each vertex has a different set of connections, which are handled efficiently using a dynamic prefix minimum query data structure. 
Such a data structure can be obtained using a data structure for reporting points in a rectangular region of the plane~\cite{Wilkinson}.

Our oracle for directed planar graphs (Section~\ref{sec::directed}) is based on the static vertex-labeled distance oracle of~\cite{MozesS15}, which uses connection for sets of vertices (i.e., a label) rather than connections for individual vertices. We show how to efficiently maintain the connections for a dynamically changing set of vertices using a bottom-up approach along the decomposition of the graph.

Our data structures support both queries and updates in polylogarithmic time. No previously known data structure supported both queries and updates in sublinear time. The following table summarizes the comparison between our oracles and the relevant previously known ones.

\begin{table}[h]
\centering
\caption{Vertex-to-Label Distance Oracles time bound comparison}
\label{tbl::oracles_time_comparison}
\begin{tabular}{|l|l|l|l|}
\hline
                  							& D/U	& Query time								& Update time					 		\\ \hhline{|=|=|=|=|}
Li, Ma and Ning~\cite{LMN13}   				& U 		& $O(\epsm\log{n}\log\Delta)$    					& $O(n\log{n})$    							\\ \hline
Łącki at el.~\cite{Lacki}      				& U 		& $O(\epsm\sqrt{n}\log^2{n}\log D)$				& $O(\epsm\sqrt{n}\log^2{n}\log D)$  				\\ \hline
Section~\ref{sec::fast_query} (faster query)		& U 		& $O(\epsm \log{n}\lglg{nN})$ 					& $O(\epsm \log{n}\lglg{n}\log{nN})$ 			\\ \hline
Section~\ref{sec::fast_update} (faster update)	& U 		& $O(\epsm\frac{\log^2{(\epsm n)}}{\lglg(\epsm n)})$	& $O(\epsm\log^{1.51}(\epsm n))$ 	\\ \hline 
Mozes and Skop~\cite{MozesS15}    			& D   		& $O(\epsm+\lglg{n}\lglg{nN})$    				& N/A    									\\ \hline
Section~\ref{sec::directed} 				& D   		& $O(\epsm \log{n}\lglg{nN})$					& $O(\epsm\log^{3}{n}\log{nN})$ 				\\ \hline
\end{tabular}
\\
In the table above, D/U stands for Directed and Undirected graphs.
\end{table}

\ifdefined \fullver \else {
	For brevity we defer all formal proofs in this paper to the Appendix.
} \fi

\mysection{Preliminaries}\label{sec::preliminaries}
We shall use the term edges and arcs when referring to undirected and directed graphs, respectively.
Given an undirected graph $G$ with a spanning tree $T$ rooted at $r$ and an edge $uv$ not in $T$,  the {\em fundamental cycle} of $uv$ (with respect to $T$) is the cycle composed of the $r$-to-$u$ and $r$-to-$v$ paths in $T$, and the edge $uv$.
By a spanning tree of a \emph{directed graph} $G$ we mean a spanning tree of the underlying undirected graph of $G$.

Let $\ell:~E(G)\rightarrow\mathbb{R}$ be a non-negative length function. Let $N$ be the ratio of the maximum and minimum values of $\ell(\cdot)$. 
We assume, for ease of presentation, that shortest paths are unique. Let $\delta_G(u,v)$ denote the $u$-to-$v$ distance in $G$ (w.r.t. $\ell(\cdot)$). 

For a simple path $Q$ and a vertex set $U\subseteq V(Q)$ with $|U|\geq 2$, we define $Q_U$, the \emph{reduction} of $Q$ to $U$ as a path whose vertices are $U$. Consider the vertices of $U$ in the order in which they appear in $Q$. For
every two consecutive vertices $u_1, u_2$ of $U$ in this order, there is an arc $u_1 u_2$ in $Q_U$ whose length is the length of the $u_1$-to-$u_2$ sub-path of $Q$.

Let $\mathcal{L}$ be a set of labels. We say that a graph $G$ is \emph{vertex-labeled} if every vertex is assigned a single label from $\mathcal L$.
For a label $\lambda\in\mathcal{L}$, let $S_G^\lambda$ denote the set of vertices in $G$ with label $\lambda$. We define the distance from a vertex $u \in V(G)$ to the label $\lambda$ by $\delta_G(u,\lambda)=\min_{v\in S_G^\lambda}\delta_G(u,v)$. If $G$ does not contain the label $\lambda$, or $\lambda$ is unreachable from $u$, we say that $\delta_G(u,\lambda)=\infty$.

\begin{definition}
For a fixed parameter $\eps\geq 0$, a \emph{stretch-$(1+\eps)$  vertex-labeled distance oracle} is a data structure that, given a vertex $u\in V(G)$ and a label $\lambda\in\mathcal{L}$, returns a distance $d$ satisfying $\delta_G(u,\lambda) \leq d \leq (1+\eps)\delta_G(u,\lambda)$. 
\end{definition}
\begin{definition}
For fixed parameters $\alpha, \eps\geq 0$,
a \emph{scale-$(\alpha, \epsilon)$ vertex-labeled distance oracle} is a data structure that, given a vertex $u\in V(G)$ and a label $\lambda\in\mathcal{L}$, such that $\delta_G(u,\lambda) \leq \alpha$, returns a distance $d$ satisfying $\delta_G(u,\lambda) \leq d \leq \delta_G(u,\lambda)+\eps\alpha$. If $\delta_G(u,\lambda) > \alpha$, the oracle returns $\infty$.	
\end{definition}

The only properties of planar graphs that we use in this paper are the existence of shortest path separators (see below), and the fact that single source shortest paths can be computed in $O(n)$ time in a planar graph with $n$ vertices~\cite{HKRS97}.

\begin{definition} Let $G$ be a directed graph. Let $G'$ be the undirected graph induced by $G$. Let $P$ be a path in $G'$. Let $S$ be a set of vertex disjoint directed shortest paths in $G$. 
	We say that $P$ is \emph{composed of} $S$ if (the undirected path corresponding to) each shortest path in $S$ is a subpath of $P$ and each vertex of $P$ is in some shortest path in $S$.
\end{definition}

\begin{definition}
Let $G$ be a directed embedded planar graph. 
An undirected cycle $C$ is a {\em balanced cycle separator} of $G$ if each of the strict interior and the strict exterior of $C$ contains at most $2|V(G)|/3$ vertices. 
If, additionally, $C$ is composed of a constant number of directed shortest paths, then $C$ is called a {\em shortest path separator}.	
\end{definition}

Let $G$ be a planar graph. We assume that $G$ is triangulated since we can triangulate $G$ with infinite length edges, so that distances are not affected.  It is well known~\cite{LiptonT79,Thorup04} that for any spanning tree of $G$, there exists a fundamental cycle $C$ that is a cycle separator. Such a cycle can be found in linear time. 
Note that, if $T$ is chosen to be a shortest path tree, or if any root-to-leaf path of $T$ is composed of a constant number of shortest paths, then the fundamental cycle $C$ is a shortest path separator. 

\mysubsection[Existing techniques]{Existing techniques for approximate distance oracles for planar graphs}

Thorup shows that to obtain a stretch-$(1+\eps)$ distance oracle, it suffices to show scale-$(\alpha,\eps)$ oracles for so-called \emph{$\alpha$-layered} graphs. An $\alpha$-layered graph is one equipped with a spanning tree $T$ such that each root-to-leaf path in $T$ is composed of $O(1)$ shortest paths, each of length at most $\alpha$. 
This is summarized in the following lemma:
\begin{lemma}~\cite[Lemma 3.9]{Thorup04}
\label{thorup::lem::stretch_eps}
For any planar graph $G$ and fixed parameter $\eps$, a stretch-$(1+\eps)$ distance oracle can be constructed using $O(\log{nN})$ scale-$(\alpha,\eps')$ distance oracles for $\alpha$-layered graphs, where $\alpha=2^i$, $i=0,...\lceil{\log{nN}}\rceil$ and $\eps'\in\{1/2, \eps/4\}$.
If the scale-$(\alpha,\eps')$ has query time $t(\epsilon')$ independent of $\alpha$, the stretch-$(1+\eps)$ distance oracle can answer queries in $O(t(1/2)\epsm+t(\eps/4)\lglg(nN))$.
\end{lemma}

All of our distance oracles are based on a recursive decomposition of $G$ using shortest path separators.
If $G$ is undirected (but not necessarily $\alpha$-layered), we can use any shortest path tree to find a shortest path separator in linear time. 
Similarly, if $G$ is $\alpha$-layered, we can use the spanning tree $G$ is equipped with to find a shortest path separator in linear time. 

We recursively decompose $G$ into subgraphs using shortest path separators until each subgraph has a constant number of vertices. We represent this decomposition by a binary tree $\mathcal{T}_G$. To distinguish the vertices of $\mathcal{T}_G$ from the vertices of $G$ we refer the former as \emph{nodes}.

Each node $r$ of $\mathcal{T}_G$ is associated with a subgraph $G_r$. The root of $\mathcal T_G$ is associated with the entire graph $G$. We sometimes abuse notation and equate nodes of $\mathcal T_G$ with their associated subgraphs. For each non-leaf node $r\in \mathcal T_G$, let $C_r$ be the shortest path separator of $G_r$. Let  \sep{r} be the set of shortest paths $C_r$ is composed of. The subgraphs $G_{r_1}$ and $G_{r_2}$ associated with the two children of $r$ in $\mathcal T_G$ are the interior and exterior of $C_r$ (w.r.t. $G_r$), respectively. Note that $C_r$ belongs to both $G_{r_1}$ and $G_{r_2}$.  For a vertex $v\in V(G)$, we denote by $r_v$ the leaf node of $\mathcal T_G$ that contains $v${\ifdefined \fullver (See figure \ref{fig:decomposition} for illustration)\fi}.

\ifdefined \fullver 
\begin{figure}
\begin{center}
\ifdefined \isarticle 
	\includegraphics[width=0.5\textwidth]{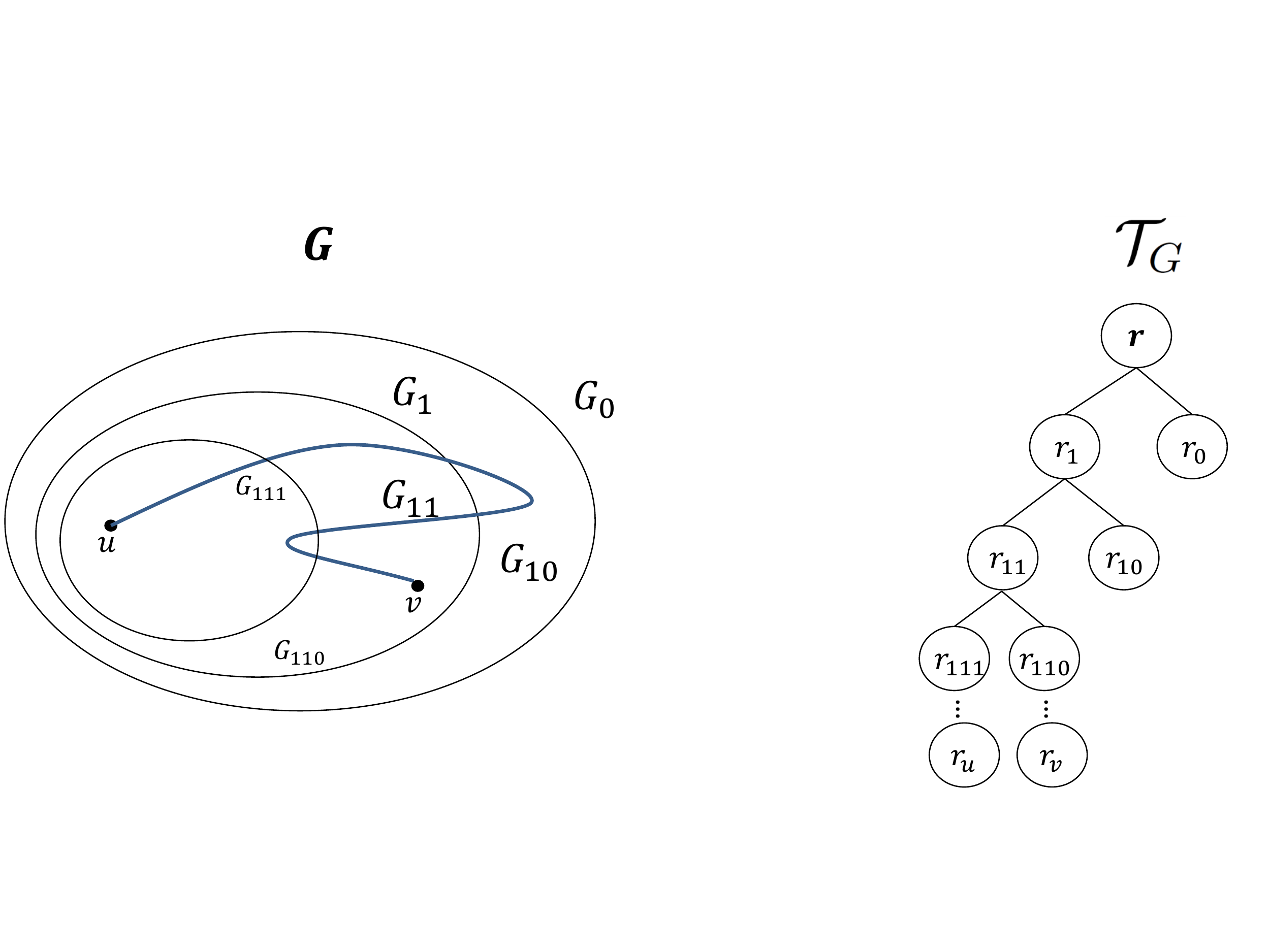}
\else
	\includegraphics[width=0.7\textwidth]{graph_decomp}
\fi
\end{center}
\label{fig:decomposition}
\caption{
An illustration of (part of) the recursive decomposition of a graph $G$ using cycle separators, and the corresponding decomposition tree $\mathcal T_G$.
The graph $G$ is decomposed using a cycle separator into $G_0$, and $G_1$. Similarly, $G_1$ is decomposed into $G_{10}$ and $G_{11}$, and $G_{11}$ is decomposed into $G_{110}$ and $G_{111}$.  The node $r$ is the root of $\mathcal T_G$ and is associated with $G_r = G$.  Similarly, $r_1$ is associated with $G_1$, etc. The nodes $r_u$ and $r_v$ are the leaf nodes that contains $u$ and $v$, respectively. The node $r_1$ is the root-most node that is intersected by the shortest $u$-to-$v$ path in $G$ (marked in blue), hence, the path is fully contained in $G_{r_1}$.
}
\end{figure}
\fi

We now describe the basic building block used in our (and in many previous) distance oracle.
Let $u,v$ be vertices in $G$. Let $Q$ be a path on the root-most separator (i.e., the separator in the node of $\mathcal T_G$ closest to its root) that is intersected by the shortest $u$-to-$v$ path $P$. 
Let $t$ be a vertex in $Q\cap P$. Note that $\delta_G(u,v)=\delta_G(u,t)+\delta_G(t,v)$.
Therefore, if we stored for $u$ the distance to every vertex on $Q$, and for $v$ the distance from every vertex on $Q$, we would be able to find $\delta_G(u,v)$ by iterating over the vertices of $Q$, and finding 
the one minimizing the distance above. This, however, is not feasible since the number of vertices on $Q$ might be $\theta(|V(G)|)$. Instead, we store the distances for a subset of $Q$. This set is called an {\em $(\alpha, \eps)$-covering connections set}. 

\begin{definition}[$(\alpha,\eps)$-covering connections set]~\cite[Section 3.2.1]{Thorup04}\label{def::eps_cover}
Let $\eps,\alpha \geq 0$ be fixed constants. Let $G$ be a directed graph. Let $Q$ be a shortest path in $G$ of length at most $\alpha$. For $u\in V(G)$ we say that $C_{G}(u,Q)\subseteq V(Q)$ is an {\em $(\alpha, \eps)$-covering connections set} from $u$ to $Q$ if and only if for every vertex $t$ on $Q$ s.t. $\delta_G(u,t) \leq \alpha$, there exists a vertex $q\in C_{G}(u,Q)$ such that $\delta_G(u,q)+\delta_G(q,t)\leq\delta_G(u,t)+\eps\alpha$.
\end{definition}
One defines $(\alpha, \eps)$-covering connections sets $C_{G}(Q,u)$ from $Q$ to $u$ symmetrically. 
Thorup proves that there always exists an $(\alpha, \eps)$-covering connections set of size $O(\epsm)$:
\begin{lemma}~\cite[Lemma 3.4]{Thorup04}\label{directed::thm::sets_size} Let $G,Q,\eps,\alpha$ and $u$ be as in definition~\ref{def::eps_cover}. There exists an $(\alpha, \eps)$-covering connections set $C_G(u,Q)$ of size at most $\lceil{2\epsm}\rceil$. 
This set can be found in $O(|Q|)$ if the distance from $u$ to every vertex on $Q$ is given.
\end{lemma}
We will use the term $\eps$-covering connections set whenever $\alpha$ is obvious from the context.
Thorup shows that $(\alpha, \eps)$-covering connections sets can be computed efficiently.
\begin{lemma}
\cite[Lemma 3.15]{Thorup04}\label{thorup::construction_time}
Let $H$ be an $\alpha$-layered graph. In $O(\epsmm{n}\log^3{n})$ time and $O(\epsm n\log{n})$ space one can compute and store a decomposition $\mathcal T_H$ of $H$ using shortest path separators, along with $(\alpha, \eps)$-covering connections sets $C_H(u,Q)$ and $C_H(Q,u)$ for every vertex $u\in V(H)$, every ancestor node $r$ of $r_u$ in $\mathcal T_H$, and every $Q\in Sep_r$.
\end{lemma}

\mysection[Undirected Graphs With Faster Query]{Oracle for Undirected Graphs With Faster Query}\label{sec::fast_query}
Let $H$ be an undirected $\alpha$-layered graph,\footnote{
The discussion of $\alpha$-layered graphs in Section \ref{sec::preliminaries} refers to directed graphs, and hence also applies to undirected graphs.} and let $T$ be the associated spanning tree of $H$. For any fixed parameter $\eps'$ we set $\eps=\frac{\eps'}{3}$. We decompose $H$ using shortest path separators w.r.t. $T$. Let $\mathcal{T}_H$ be the resulting decomposition tree.
For every node $r\in \mathcal{T}_H$ and every shortest path $Q\in Sep_r$, we select a set $C_Q\subseteq V(Q)$ of $\epsm$ connections evenly spread intervals along $Q$\footnote{We assume that the endpoints of the intervals are vertices on $Q$, since otherwise once can add artificial vertices on $Q$ without asymptotically change in the size of the graph.}.
Thus, for every vertex $t\in V(Q)$ there is a vertex $q\in C_Q$ such that $\delta_H(t,q)\leq\eps\alpha$.

We compute in $O(|H_r|)$ time a shortest path tree in $H_r$ rooted at each $q\in C_q$ using~\cite{HKRS97}. This computes for every $u\in V(H)$, the connection length 
$\delta_{H_r}(u,q)$.

\begin{restatable}{lemma}{constconnectionset}
\label{const_connection_set}
Let $u\in V(H)$. For every ancestor node $r\in \mathcal{T}_H$ of $r_u$, and every $Q\in Sep_r$, $C_Q$ is a $2\eps$-covering connections set from $u$ to $Q$.
\end{restatable}
\defproof{constconnectionset}{
\begin{proof}
Let $t\in Q$. We need to show that there exist $q\in C_Q$ such that $\delta_{H_r}(u,t)\leq\delta_{H_r}(u,q)+\delta_{H_r}(q,t)\leq \delta_{H_r}(u,t)+\eps'\alpha$. Since $t\in Q$, there exists a vertex $q\in C_Q$ such that $\delta_H(q,t)\leq\eps\alpha$.
Since $H$ is undirected, the triangle inequality for shortest path lengths holds for any three vertices in $V(H)$. We start with the triangle inequality between $u$, $t$ and $q$ in $H$ as follows.
\begin{align*}
\delta_{H_r}(u,q)&\leq\delta_{H_r}(u,t) + \delta_{H_r}(t,q)\\
\delta_{H_r}(u,q)+ \delta_{H_r}(t,q)&\leq\delta_{H_r}(u,t) + \delta_{H_r}(t,q)+ \delta_{H_r}(t,q)\\
\delta_{H_r}(u,q)+ \delta_{H_r}(t,q)&\leq\delta_{H_r}(u,t) + 2\eps\alpha
\end{align*}

From the triangle inequality, $\delta_{H_r}(u,t)\leq\delta_{H_r}(u,q)+\delta_{H_r}(q,t)$, and the lemma follows.
\qendproof}

\ifdefined \fullver 
\begin{figure}[h]
\begin{center}
\ifdefined \isarticle 
	\includegraphics[width=0.5\textwidth]{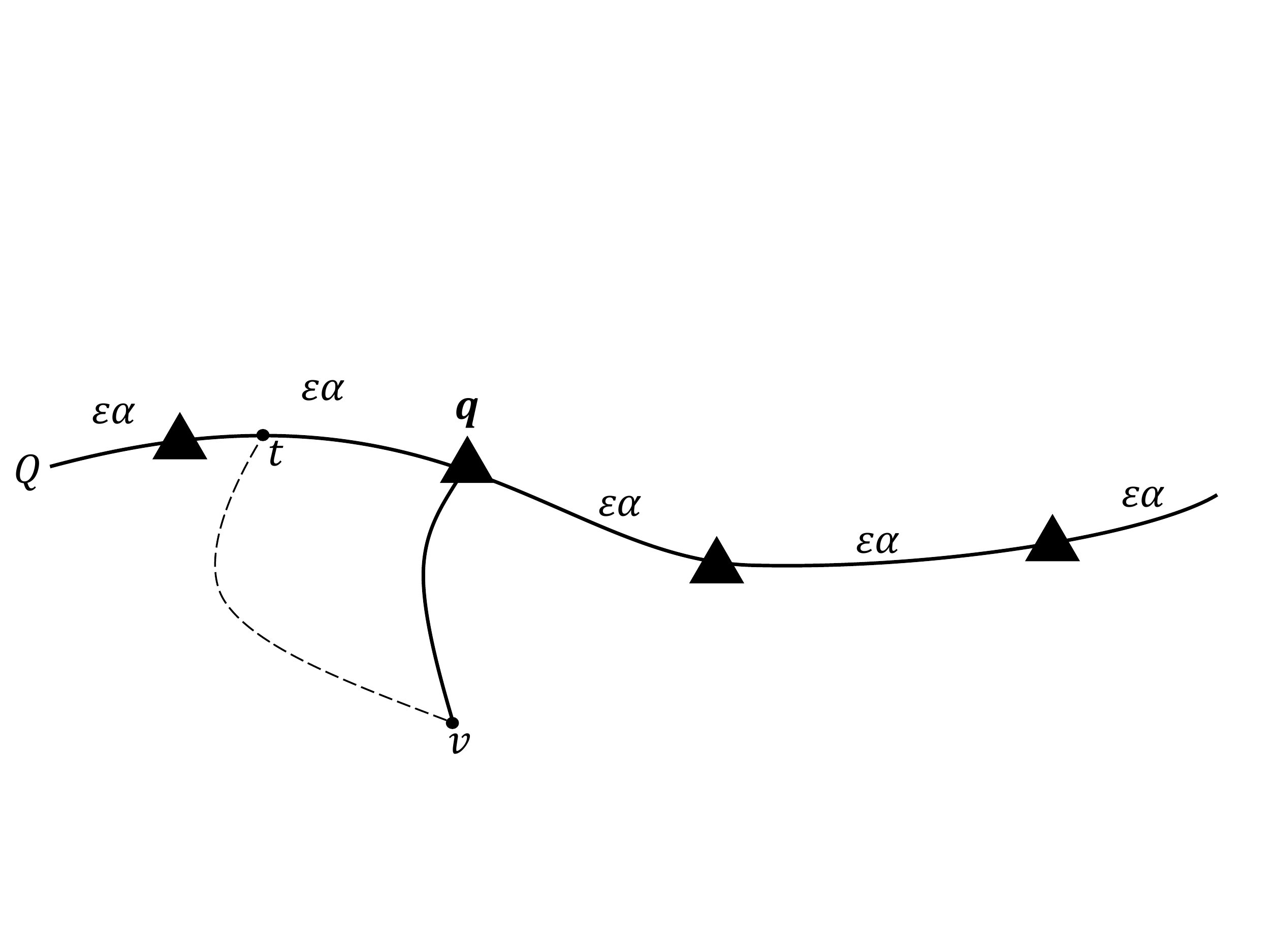}
\else
	\includegraphics[width=0.7\textwidth]{evenly_spaced}
\fi
\end{center}
\label{fig:evenly_spaced}
\caption{
Illustration of Lemma~\ref{const_connection_set}.
$Q$ is a shortest path in some separator, the connections of $C_Q$ are marked by triangles. The solid $v$-to-$q$ path reflects the shortest path from $v$ to the connection $q$, and the dashed $v$-to-$t$ path reflects the shortest path from $v$ to $t$.
}
\end{figure}
\fi

\mysubsection{Warm up: the static case}
We start by describing our data structure for the static case with a single fixed label $\lambda$. 
For every node $r\in \mathcal{T}_H$, let $S_{r}^{\lambda}$ be the set of $\lambda$-labeled vertices in $V(H_r)$. For every separator $Q\in Sep_r$, every vertex $q\in C_Q$, and every vertex $v\in S_{r}^{\lambda}$ let $\hat\delta_{H_r}(q,v)=k\eps\alpha$ where $k$ is the smallest value such that $\delta_{H_r}(q,v)\leq k \eps\alpha$. Thus, $\delta_{H_r}(q,v)\leq\hat\delta_{H_r}(q,v)\leq\delta_{H_r}(q,v)+\eps\alpha$.
Let $L_r(q,\lambda)$ be the list of the distances $\hat\delta_{H_r}(q,v)$ for all $v\in S_r^\lambda$.
We sort each list in ascending order. Thus, the first element of $L_r(q,\lambda)$ denoted by $first(L_r(q,\lambda))$ is at most $\eps\alpha$ more than the distance from $q$ to the closest $\lambda$-labeled vertex in $H_r$.
We note that each vertex $u\in V(H)$ may contribute its distance to $O(\epsm \log{n})$ lists. Hence, we have $O(\epsm n\log{n})$ elements in total. Since $H$ is an $\alpha$-layered graph, the length of $Q$ is bounded by $\alpha$. Hence, the universe of these lists is bounded by $\frac{\alpha}{\eps\alpha}=\epsm$. Thus, these lists can be sorted in total  $O(\epsm n\log{n})$ time.

\mysubsubsection{Query($u$,$\lambda$)} Given $u\in H$.
We wish to find the closest $\lambda$-labeled vertex $v$ to $u$ in $H$.
For each ancestor $r$ of $r_u$, for each $Q\in Sep_r$, we perform the following search.
We inspect for every $q\in C_Q$, the distance $\delta_{H_r}(u,q)+first(L_r(q,\lambda))$. We also inspect the $\lambda$-labeled vertices in $H_{r_u}$ explicitly.
We return the minimum distance inspected.

\ifdefined \fullver 
\begin{figure}
\begin{center}
\ifdefined \isarticle 
	\includegraphics[width=0.5\textwidth]{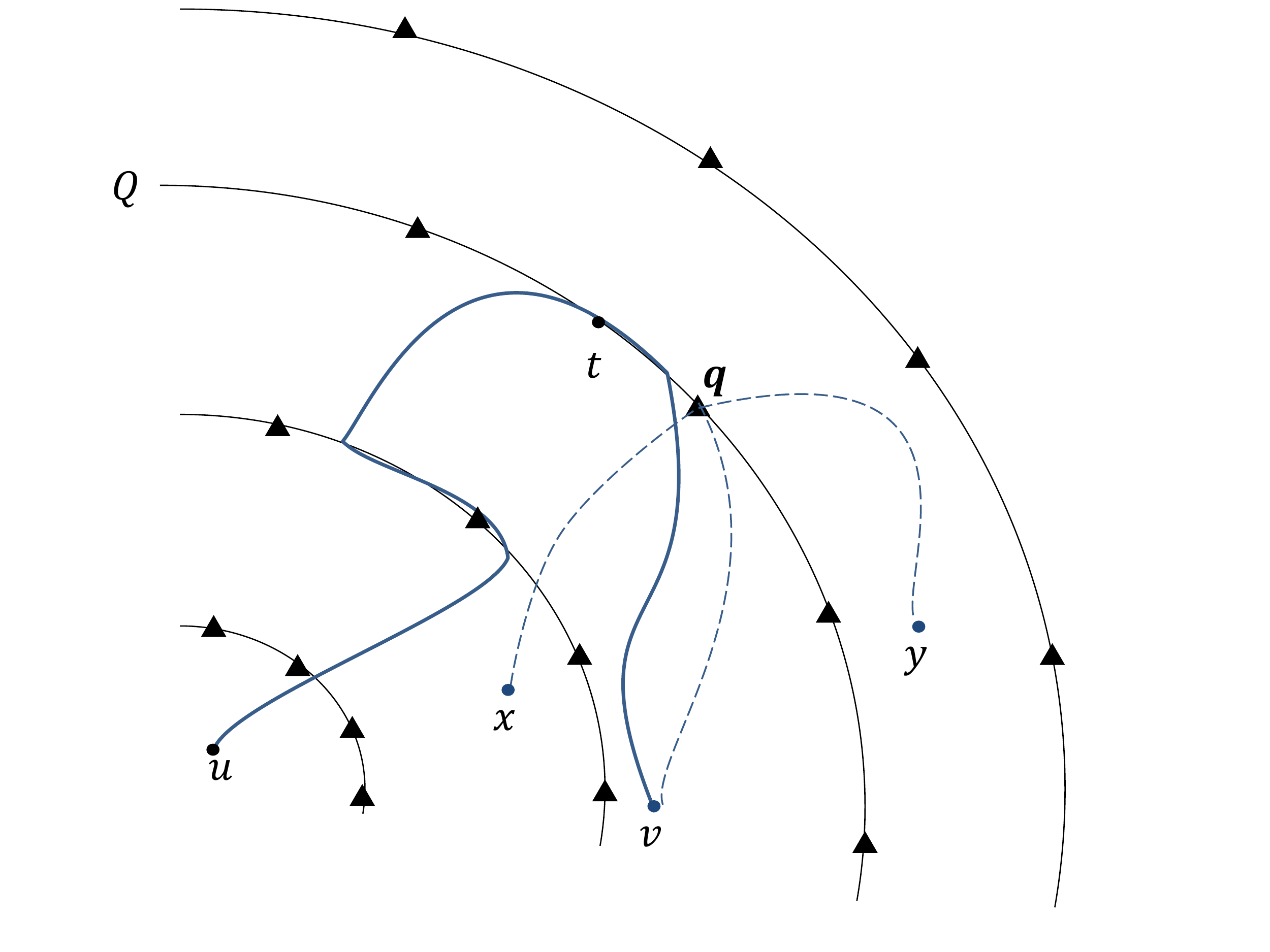}
\else
	\includegraphics[width=0.7\textwidth]{fast_query_alg}
\fi
\end{center}
\label{fig:fast_query_alg}
\caption{
Illustration of the query algorithm. The solid quarter-circles  are shortest paths of separators in $G$. The vertices $x$, $y$ and $v$ have label $\lambda$, and $v$ is the closest $\lambda$-labeled vertex to $u$. The path $Q$ belongs to the root-most node $r$ whose separator is intersected by the shortest $u$-to-$\lambda$ path (solid blue). The vertices $q$ and $t$ on $Q$ are as in the proof of Lemma~\ref{lem::fast_query::correct}. The connection $q$ minimizes $\delta_H(u,q)$ + $first(L_r(q,\lambda))$. The distances in $L_r(q,\lambda)$ are the lengths of the dashed paths.
}
\end{figure}
\fi

\begin{restatable}{lemma}{fastquerycorrect}
\label{lem::fast_query::correct}
The query algorithm runs in $O(\epsm \log{n})$ time, and returns a distance $d$ such that $\delta_H(u,\lambda)\leq d \leq \delta_H(u,\lambda)+3\eps\alpha$.
\end{restatable}

\defproof{fastquerycorrect}{
\begin{proof}
Let $v$ be the closest $\lambda$-labeled to $u$ in $H$. It is trivial that if the shortest path $P$ form $u$-to-$v$ does not leave $r_u = r_v$ the query algorithm is correct, since the distances in $r_u$ are computed explicitly. Otherwise, let $r$ be the root-most node in $\mathcal{T}_H$ such that $P$ intersects some $Q\in Sep_r$. Thus, $P$ is fully contained in $H_r$.
Let $t$ be a vertex in $Q\cap P$. Since $v$ is the closest $\lambda$-labeled vertex to $u$, it follows that it is also the closest $\lambda$-labeled vertex to $t$. 

Since $t\in Q$, there exists $q\in C_Q$ such that $\delta_{H_r}(v,q)+\delta_{H_r}(q,t)\leq\delta_{H_r}(v,t)+\eps'\alpha$. By the triangle, $\delta_{H_r}(v,q)\leq\delta_{H_r}(q,t)+\delta_{H_r}(v,t)$. Hence, $first(L_r(q,\lambda))\leq\delta_{H_r}(q,t)+\delta_{H_r}(v,t)\leq\delta_{H_r}(q,v)+\eps\alpha$. 
\begin{align}
first(L_r(q,\lambda)&\leq\hat\delta_{H_r}(q,v)\leq\delta_{H_r}(q,v)+\eps\alpha \label{undirected::fast_query::eq_1}\\
&\leq\delta_{H_r}(q,t)+\delta_{H_r}(t,v)+\eps\alpha \label{undirected::fast_query::eq_2}\\
&\leq\delta_{H_r}(t,v)+2\eps\alpha \label{undirected::fast_query::eq_3}
\end{align}
Where inequality (\ref{undirected::fast_query::eq_1}) follows from the definition of $L_r(q,\lambda)$, (\ref{undirected::fast_query::eq_2}) follows from the triangle inequality, and (\ref{undirected::fast_query::eq_3}) follows from the fact that $\delta_{H_r}(q,t)\leq \eps\alpha$.
\begin{align}
Query(u,\lambda)\leq&\delta_{H_r}(u,q)+ first(L_r(q,\lambda))\\
	\leq&\delta_{H_r}(u,q)+\delta_{H_r}(t,v)+2\eps\alpha\\
	\leq&\delta_{H_r}(u,t)+\delta_{H_r}(t,q)+\delta_{H_r}(t,v)+2\eps\alpha \label{undirected::fast_query::eq_4}\\
	\leq&\delta_{H_r}(u,v)+3\eps\alpha\\
	\leq&\delta_H(u,\lambda)+3\eps\alpha		\label{undirected::fast_query::eq_5}
\end{align}
Here, inequality (\ref{undirected::fast_query::eq_4}) follows from the triangle ineqaulity, and (\ref{undirected::fast_query::eq_5}) follows from the fact that $P$ is fully contained in $H_r$, and our assumption that $v$ is the closest $\lambda$-labeled vertex to $v$.

Since $\delta_{H_r}(u,q)+ first(L_r(q,\lambda))$ underlines a real path in the $H_r$, from our assumption that $v$ is the closest $\lambda$-labeled vertex to $u$, it follows that $Query(u,\lambda)\geq\delta_{H_r}(u,v)$, and the lemma follows.

To prove the query time, observe that 
the height of $\mathcal{T}_H$ is $O(\log{n})$. At any level of the decomposition we inspect the first element in $O(\epsm)$ lists, that is $O(\epsm \log{n})$ time. We also inspect constant number of distances in $r_u$ in constant time.
\qendproof
}

We now generalize to multiple labels. Let $\mathcal{L}$ be the set of labels in $H$. For $r\in \mathcal{T}_H$, let $\mathcal{L}_r$ be the restriction of $\mathcal{L}$ to labels that appear in $H_r$. For every label $\lambda\in \mathcal{L}_r$, every $Q \in Sep_r$ and every $q \in C_Q$, we store the list $L_r(q,\lambda)$. This does not affect the total size of our structure, since each vertex has one label, so it still contributes its distances to $O(\epsm \log{n})$ lists. The proof of Lemma~\ref{lem::fast_query::correct} remains the same since each list contains distances to a single label. 

Naively, we could store for every node $r$, every vertex $q$, and every label $\lambda\in\mathcal{L}$ the list $L_r(q,\lambda)$ in a fixed array of size $|\mathcal{L}|$. This allows $O(1)$-time access to each list, but increases the space by a factor of $|\mathcal{L}|$ w.r.t. the single label case.
Instead, we use hashing. Each vertex $q$ holds a hash table of the labels that contributed distances to $q$. For the static case, one can use perfect hashing~\cite{FKS} with expected construction time and constant query time. In the dynamic case, we will use a dynamic hashing scheme, e.g.,~\cite{PaghR01}, which provides query and deletions in $O(1)$ worst case, and insertions in $O(1)$ expected amortized time.

\mysubsection{The dynamic case}
We now turn our attention to the dynamic case. We wish to use the following method for updating our structure.
When a node $v$ changes its label from $\lambda_1$ to $\lambda_2$, we would like to iterate over all ancestors $r$ of $r_v$ in $\mathcal{T}_H$.
For every $Q\in Sep_r$ and every $q\in C_Q$, we wish to remove the value contributed by $v$ from $L_r(q,\lambda_1)$, and insert it to $L_r(q,\lambda_2)$. We must maintain the lists sorted, but do not wish to pay $O(\log{n})$ time per insertion to do so. We will be able to pay $O(\lglg{n})$ per insertion/deletion by using a successor/predecessor data structure as follows.

For every $r\in \mathcal{T}_H$, $Q\in Sep_r$, and $q\in C_Q$, let $L_r(q)$ be the list containing {\em all distances} from all vertices in $V(H_r)$ to $q$ sorted in ascending order. We note that since the distance for each specific vertex to $q$ does not depend on its label, the list $L_r(q,\lambda)$ is a restriction of $L_r(q)$ to the $\lambda$-labeled vertices in $H_r$.

During the construction of our structure we build $L_r(q)$, and, for every vertex $v$ in $H_r$, we store for $v$ its corresponding index in $L_r(q)$. We denote this index as $ID_q(v)$. We also store for $q$ a single lookup table from the IDs to the corresponding distances. 
We note that $v$ has $O(\epsm \log n)$ such identifiers, and in total we need $O(\epsm n\log{n})$ space to store them.

Now, instead of using linked list as before, we implement $L_r(q,\lambda)$ using a successor/predecessor structure over the universe $[1,...,|V(H_r)|]$ of the IDs. For example, we can use y-fast tries~\cite{Willard83} that support operations in $O(\lglg{n})$ expected amortized time and minimum query in $O(1)$ worst case.

\mysubsubsection{$Query(u,\lambda)$}
The query algorithm remains the same as in the static case. For every ancestor $r$ of $r_u$ in $\mathcal{T}_H$, every $Q\in Sep_r$, and every connection $q\in C_Q$, we retrieve the minimal ID from $L_r(q,\lambda)$ , and use the lookup table to get the actual distance between $q$ and the vertex with that ID.

\mysubsubsection{Update}
Assume that the vertex $v$ changes its label from $\lambda_1$ to $\lambda_2$. For every ancestor $r$ of $r_v$ in $\mathcal{T}_H$, every $Q\in Sep_r$, and every $q\in C_Q$, we remove $ID_q(v)$ from $L_r(q,\lambda_1)$ and insert it to $L_r(q,\lambda_2)$.

\begin{restatable}{lemma}{constutime}
\label{const_utime}
The update time is $O(\epsm\log{n}\lglg{n})$ expected amortized.
\end{restatable}

\defproof{constutime}{
\begin{proof}
In each one of the $O(\log{n})$ levels in $\mathcal{T}_H$, we perform $O(\epsm)$  insertions and deletions from successor/predecessor structures in $O(\lglg{n})$ expected amortized time per operation. Therefore the total update time is $O(\epsm\log{n}\lglg{n})$.
If the set $\mathcal{L}_r$ changes for some $r\in\mathcal{T}_H$ as a result of the update, we must also update the hash table that handles the labels. This might cost an additional $O(1)$ expected amortized time per node, and is bounded by $O(\log{n})$ expected amortized time in total.
\qendproof}

\begin{restatable}{lemma}{constspace}
\label{const_space}
The data structure can be constructed in $O(\epsm n\log{n}\lglg{n})$ expected amortized time, and stored using $O(\epsm n\log{n})$ space.
\end{restatable}

\defproof{constspace}{
\begin{proof}
We decompose $H$ into $\mathcal{T}_H$, and compute the connection length in $O(\epsm n\log{n})$ time. 
We than build the lists $L_r(q)$ for every node $r\in \mathcal{T}_H$ and $q$ on any $q\in Sep_r$. These lists contains $O(\epsm n\log{n})$ elements in the range $[1,...,\epsm]$ that is independent of both $n$ and $\alpha$. Hence we sort the lists in $O(\epsm n\log{n})$ time. 
We than use our update process on each $v\in V(H)$ and each ancestor $r$ of $r_v$ in $O(\epsm \log{n}\lglg{n})$ expected amortized time for $v$. Hence, our construction time is $O(\epsm n\lg{n}\lglg{n})$ expected amortized.
To see our space bound, we note that every $v$ contributes a distance $O(\epsm)$ lists at every ancestor $r$ of $r_v$. Hence, there are $O(\epsm n\log{n})$ elements in total. Our successor/predecessor structures, and the hash tables has linear space in the number of elements stored. Thus, $O(\epsm n\log{n})$ space.
\qendproof
}

\noindent We plug in this structure to Lemma~\ref{thorup::lem::stretch_eps} and obtain the following theorem:\footnote{Formally, one needs to show that Lemma~\ref{thorup::lem::stretch_eps} holds for vertex-labeled oracles as well. See Appendix~\ref{apndx::scale_to_stretch_reduction}.}

\begin{theorem}\label{thm::fast_query}
Let $G$ be an undirected planar graph. There exists a stretch-$(1+\eps)$ Approximate Dynamic Vertex-Labeled Distance Oracle that supports query in $O(\epsm \log{n}\lglg{nN})$ worst case and updates in $O(\epsm \log{n}\lglg{n}\log{nN})$ expected amortized. The construction time of that oracle is $O(\epsm n\log{n}\lglg{n}\log{nN})$ and it can be stored in $O(\epsm n\log{n}\log{nN})$ space.
\end{theorem}

\mysection[Directed Graphs]{Oracle for Directed Graphs}\label{sec::directed}
For simplicity we only describe an oracle that supports queries from a given label to a vertex. Vertex to label queries can be handled symmetrically.
To describe our data structure for directed graphs, we first need to introduce the concept of $\eps$-covering set from a \emph{set of vertices} to a directed shortest path.

\begin{definition}\label{def::covering_set_of_a_set}
Let $S$ be a set of vertices in a directed graph $H$. Let $Q$ be a shortest path in $H$ of length at most $\alpha$. $C_H(S,Q)\subseteq V(Q)\times\mathbb{R}^{+}$ is an {\em $\eps$-covering set} from $S$ to $Q$ in $H$ if for every $t\in Q$ s.t. $\delta_H(S,t)\leq\alpha$, there exists $(q,\ell)\in C_H(S,Q)$  s.t. $\ell+\delta(q,t)\leq\delta_{H}(S,t)+\eps\alpha$, and $\ell\geq\delta_H(S,q)$.
\end{definition}
In the definition above we use $\ell$ instead of $\delta(S,q)$ (compare to Definition~\ref{def::eps_cover}) because we cannot afford to recompute  exact distances as $S$ changes. Instead, we store and use approximate distances $\ell$.

\begin{restatable}{lemma}{setsize}
\label{directed::lem::sets_size} Let $H$ be a directed planar graph. Let $Q$ be a shortest path in $H$ of length at most $\alpha$. For every set of vertices $S\subseteq V(H)$ there is an $\eps$-covering set $C_H(S,Q)$ of size $O(\epsm)$.
\end{restatable}

\defproof{setsize}{
\begin{proof}
We introduce a new apex vertex in $H$ denoted by $x$. For every vertex $v$ in $S$, we add an arc $xv$ with length $0$. Since the indegree of $x$ is $0$, $Q$ remains a shortest path, with length bounded by $\alpha$. We apply Lemma \ref{directed::thm::sets_size} on $x$ w.r.t $Q$, to get an $\eps$-cover set $C_H(x,Q)$ of size $O(\epsm)$. Clearly, $C_H(x,Q)$ is an $\eps$-covering set from $S$ to $Q$, and the Lemma follows.
\qendproof
}

\noindent Our construction relies on the following lemma.

\begin{restatable}[Thinning Lemma]{lemma}{thinninglem}
\label{directed::lem::thinning_lemma} 
Let $H$, $S$ and $Q$ be as in Lemma \ref{directed::lem::sets_size}. 
Let $\{S_i\}_{i=1}^k$ be sets such that $S=\bigcup_{i=1}^{k}{S_i}$. For $1 \leq i\leq k$, let $D_H(S_i,Q)$ be an $\eps'$-covering set from $S_i$ to $Q$, ordered by the order of the vertices on $Q$. Then for every $\eps > 0$, 
an $(\eps+\eps')$-covering set $C_H(S,Q)$ from $S$ to $Q$ of size $\lceil2\epsm\rceil$ can be found in $O(\epsm+|\bigcup_{i=1}^{k}{D_H(S_i,Q)}|)$ time.
\end{restatable}

\defproof{thinninglem}{
\begin{proof}

Let $q_0$ be the first vertex on $Q$. Let $\hat Q$ be the reduction of $Q$ to the vertices in $\bigcup_{i=1}^{k}{D_Q(S_i)}$  and $q_0$.
Let $\hat{H}$ be the auxiliary graph consisting of $\hat{Q}$ and an apex vertex $x$ connected to every $q\in\hat{Q}$ with an arc $xq$ of length $\delta_{H}(S_i,q)$, where $S_i$ is the set originally containing $q$. Note that $\delta_{H}(S_i,q)\geq\delta_{H}(S,q)$. 
Also note that $\hat{H}$ is planar, with diameter bounded by $\alpha$, and since the indegree of $x$ is $0$, $Q$ is a shortest path in $\hat{H}$. Let $m=|\bigcup_{i=1}^{k}{D_H(S_i,Q)}|$. We compute the shortest distance from $x$ to every other $q$ in $\hat{H}$ explicitly by relaxing all arcs adjacent to $x$, and than relaxing the arcs of $Q$ by order.  Constructing $\hat{H}$ and computing these distances can be done in $O(m)$ time, since $|V(\hat{H})|=|E(\hat{H})|=O(m)$.

We apply Lemma \ref{directed::thm::sets_size} to $x$ with $\eps$ and get an $\eps$-covering set $C_{\hat{H}}(x,Q)$ of size $\lceil2\epsm\rceil$ from $x$ to $\hat{Q}$. It remains to prove that $C_{\hat{H}}(x,Q)$ is an $(\eps+\eps')$-covering set $C_H(S,Q)$ set from $S$ to $Q$ in $H$.

Let $t\in Q$. We show that there exists $(q,\ell)\in C_{\hat{H}}(x,Q)$ such that $\ell+\delta_H(q,t)\leq\delta_H(S, t)+(\eps'+\eps)\alpha$. We assume without loss of generality, that $\delta_H(S, t)=\delta_H(S_1, t)$. Since $D_Q(S_1)$ is an $\eps'$-covering set from $S_1$ to $Q$ in $H$, there exists $(q',\ell')\in D_Q(S_1)$ such that:
\begin{equation}\label{directed::equ::1}
\ell'+\delta_H(q',t)\leq\delta_H(S_1,t)+\eps' \alpha
\end{equation}
Also, since $q'\in D_Q(S_1)$, it is also on $\hat{Q}$. Therefore there exists $(q,\ell)\in C_{\hat{H}}(x,Q)$  such that:
\begin{equation}\label{directed::equ::2}
\ell+\delta_{\hat{H}}(q,q')\leq\delta_{\hat{H}}(x,q')+\eps\alpha\leq \ell'+\eps\alpha
\end{equation}
Where the last inequality follows the fact that for every $(q^{*},\ell^{*})\in D_Q(S_1)$, $\delta_{\hat{H}}(x,q^{*})\leq \ell^{*}$, and hence, $\delta_{\hat{H}}(x,q')$ is at most $\ell'$.
\begin{align}
\delta_H(S_1\cup S_2,t)+\eps' \alpha+\eps\alpha &= \delta_H(S_1,t)+\eps' \alpha+\eps\alpha 	\label{directed::equ::3}	\\ 
&\geq \ell'+\delta_H(q',t)+\eps\alpha								\label{directed::equ::4}	\\ 
&\geq \ell+\delta_{\hat{H}}(q,q') +\delta_H(q',t)					\label{directed::equ::5}	\\ 
&= \ell+\delta_{H}(q,q') +\delta_H(q',t)							\label{directed::equ::6}	\\  
&= \ell+\delta_{H}(q,t) 									\label{directed::equ::7}
\end{align}
Here, (\ref{directed::equ::4}) follows from inequality (\ref{directed::equ::1}), (\ref{directed::equ::5}) follows from inequality (\ref{directed::equ::2}).
\qendproof
}

\ifdefined \fullver 
\begin{figure}[h]
\begin{center}
\ifdefined \isarticle 
	\includegraphics[width=0.5\textwidth]{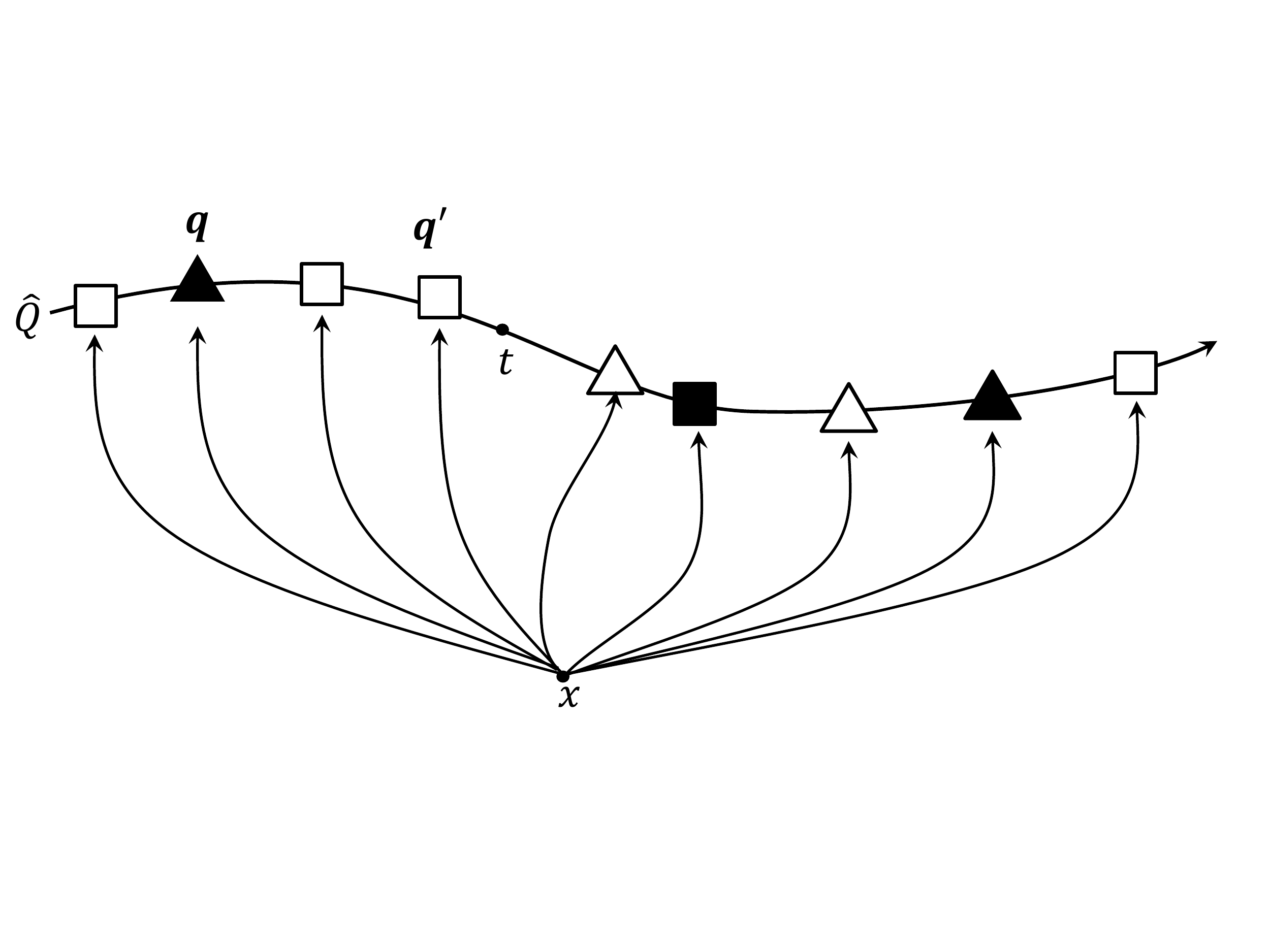}
\else
	\includegraphics[width=0.7\textwidth]{aux_graph}
\fi
\end{center}
\label{fig:aux_graph}
\caption{
Illustration of he auxiliary graph $\hat H$ in the proof of Lemma~\ref{directed::lem::thinning_lemma}, when applied to the sets $S_1 =u$ and $S_2=v$ (I.e. $k=2$). The connection sets of $S_1$ and $S_2$ are indicated by the triangles and squares, respectively. The connections in the output set are indicated by a solid fill. The vertex $t$ is a vertex on $Q$ (not on $\hat Q$) that is closer to $v$ than it is to $u$. 
Although $t$ is covered by both $q$ and $q'$, its distance from $x$ is better approximated via $q'$. The vertex $q$ $\eps$-covers $q'$ w.r.t. the distances in $\hat H$ hence $q'$ is not included in the output set. Since $q$ $\eps$-covers $q'$, and $q'$ $\eps'$-covers $t$, it follows that $t$ is $(\eps+\eps')$-covered by $q$.
}
\end{figure}
\fi

Let $H$ be a directed planar $\alpha$-layered graph, equipped with a spanning tree $T$. For every fixed parameter $\eps$, let $\hat\epsilon=\frac{\eps}{8\log{n}}$, and $\eps^{*}=\frac{\eps}{2}$. We apply Lemma \ref{thorup::construction_time} with $\hat\epsilon$ to $H$ and obtain a decomposition tree $\mathcal{T}_H$, and $\hat\epsilon$-covering sets $C_{H_r}(v,Q)$ and $C_{H_r}(Q,v)$ for every $v\in V(H)$, every ancestor $r$ of $r_v$ in $\mathcal{T}_H$ and every $Q\in Sep_r$. 
For every $1 \leq i \leq \log{n}$, let $\eps_i = \frac{\eps\log{n}- i+1}{4\log{n}}$. 

For every $r\in\mathcal T_H$, for every $\lambda\in\mathcal L_r$, and for every $Q\in Sep_r$, we apply Lemma~\ref{directed::lem::thinning_lemma} to the $\hat\eps$-covering connections sets $C_{H_r}(v,Q)$  for all $v\in S_r^\lambda$, with $\eps'=\hat\eps$ and $\eps$ set to $\frac{\eps}{4}$. Thus, we obtain an $\eps^{*}$-covering set $C^{*}_{H_r}(S^{\lambda}_{r},  Q)$.
Let $i$ be the level of $r$ in $\mathcal T_H$, we also store for $r$ a set of $\eps_i$-covering sets as follows. For every ancestor node $t$ of $r$ in $\mathcal{T}_H$ and every $Q\in Sep_t$, we store $C_{H_t}(S^{\lambda}_{r},  Q)$. We assume for the moment that these sets are given. We defer the description of their construction (see the update procedure and the proof of Lemma~\ref{directed::lem::update_time}).
We will use the $\eps^{*}$-covering sets for efficient queries, and the more accurate $\eps_i$-covering sets to be able to perform efficient updates. 

\ifdefined \fullver 
\begin{figure}
\begin{center}
\ifdefined \isarticle 
	\includegraphics[width=0.7\textwidth ]{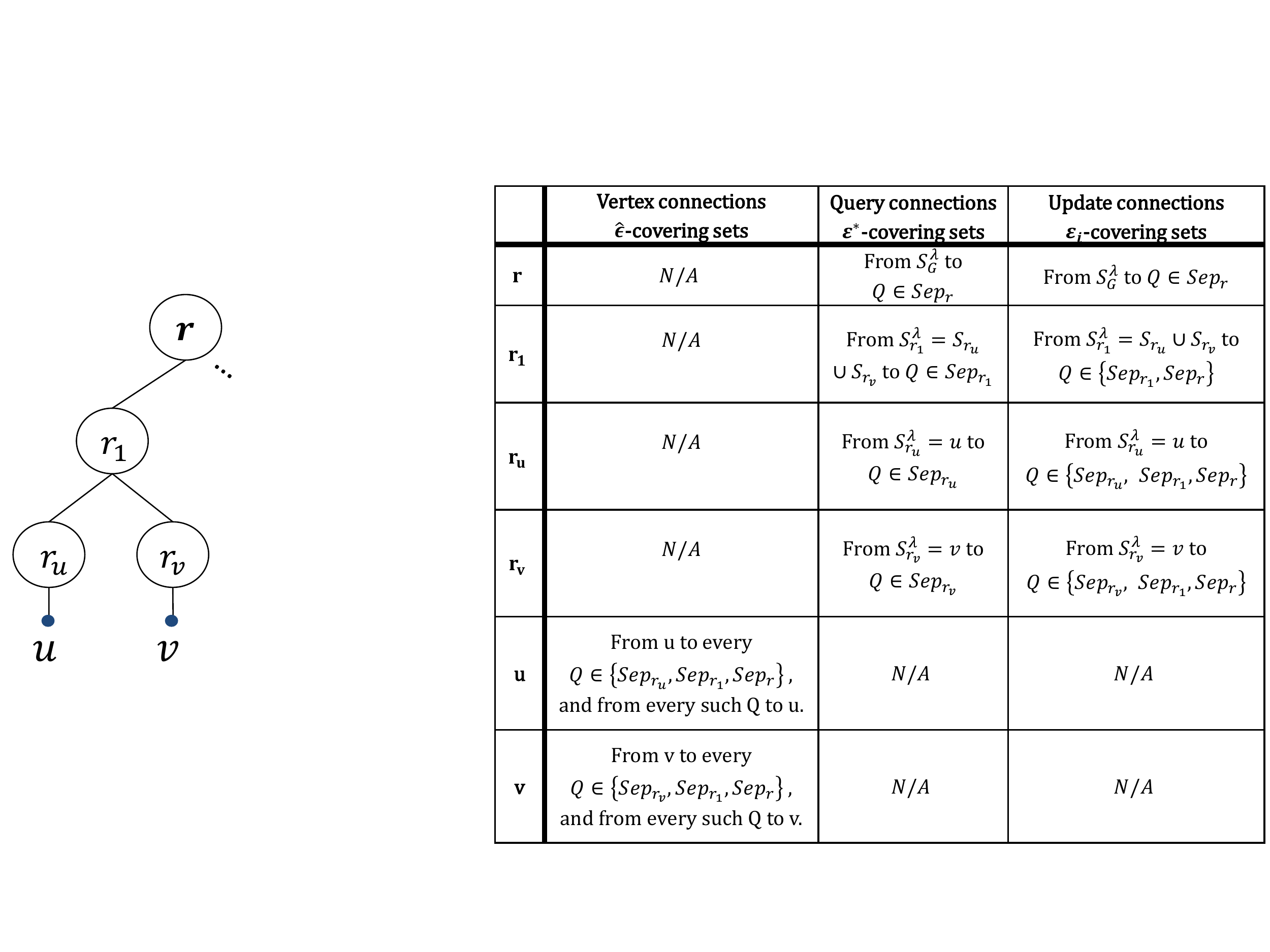}
\else
	\includegraphics[width=\textwidth ]{eps_cover_types}
\fi
\end{center}
\caption{
A summary of the connections-sets stored by the directed oracle. To the left, part of a decomposition tree of a graph. The vertices $u$ and $v$ are the only $\lambda$ labeled vertices. To the right, a table listing all the covering sets that are stored for the label $\lambda$.
}
\label{fig:eps_cover_types}
\end{figure}
\fi

\mysubsection[Query]{$Query(\lambda, u)$}
The query algorithm is straightforward. For every ancestor $r$ of $r_u$ we find $(q,\ell)\in C^{*}_{H_r}(S_{r}^{\lambda},Q)$ and $t\in C_{H_r}(Q,u)$ that minimizes the distance $\ell + \delta_{H_r}(q, t) + \delta_{H_r}(t, u)$.
We also inspect the distance to the $\lambda$-labeled vertices in $r_u$ explicitly. We return the minimum distance inspected.
To see that the query time is $O(\epsm \log{n})$, we note that for every one of the $O(\log{n})$ ancestors of $r_u$ we inspect $O(\epsm)$ distances on constant number of separators. Inspecting the distances in $r_u$ itself takes constant time.

\ifdefined \fullver 
\begin{figure}
\begin{center}
\ifdefined \isarticle 
	\includegraphics[width=0.5\textwidth]{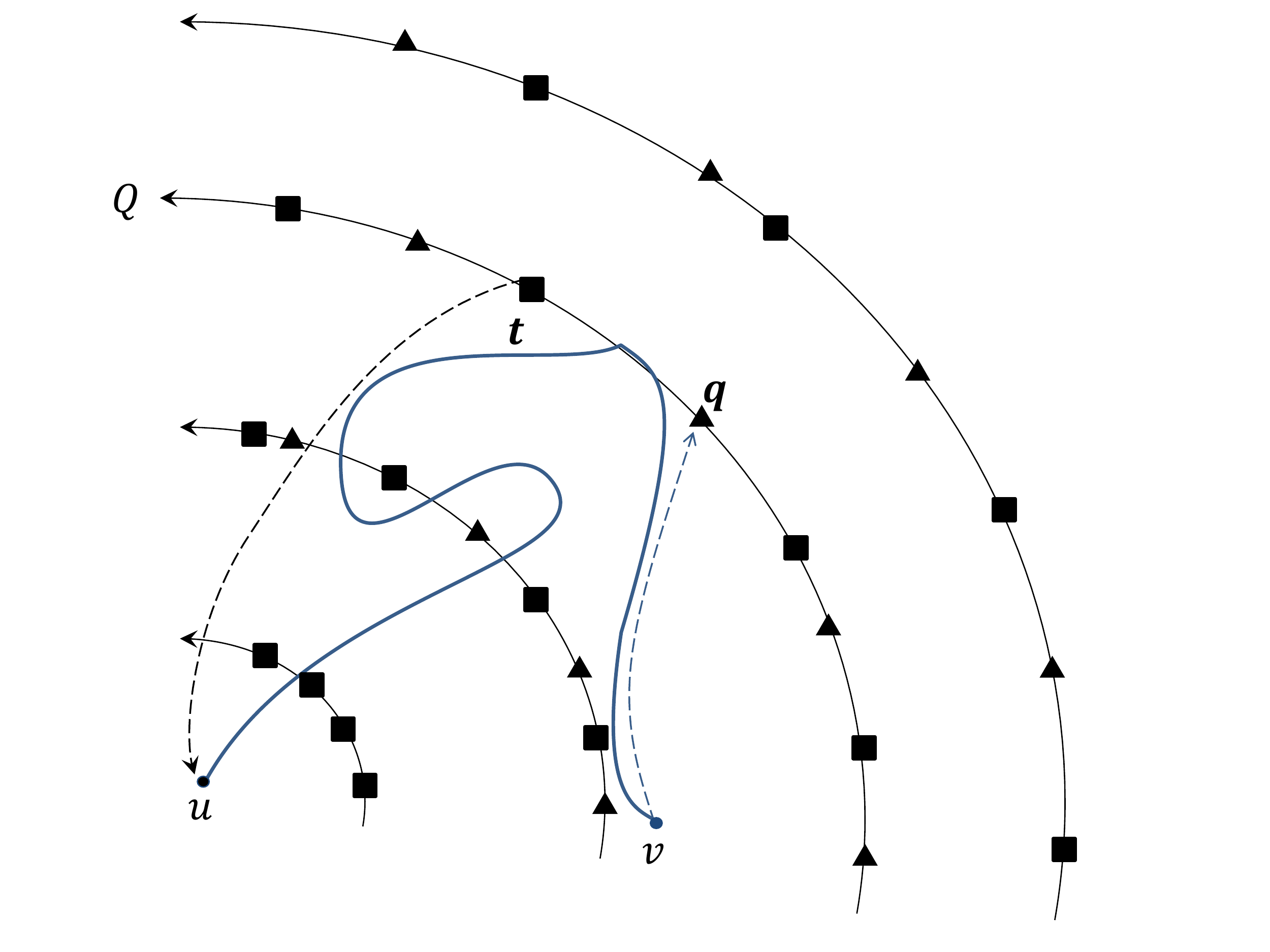}
\else
	\includegraphics[width=0.7\textwidth]{directed_query}
\fi
\end{center}
\label{fig:directed_query}
\caption{
The solid  quarter-circles are shortest paths of separators in $G$. The vertex $v$ is the closest $\lambda$-labeled vertex to $u$. The path $Q$ belongs to the root-most node $r$ whose separator is intersected by the shortest $\lambda$-to-$u$ path (solid blue). The connections of $S_r^\lambda$ on $Q$ are indicated by black triangles, the connections of $u$ are indicated by black squares. The vertices $q$ and $t$ on $Q$ are as in the proof of Lemma~\ref{directed::lem::correctness}.
The blue and black dashed lines are the shortest paths from $v$ to $q$, and from $t$ to $u$, respectively. These paths are used to generate the distance reported by the query algorithm.
}
\end{figure}
\fi

\begin{restatable}{lemma}{dircorrect}
\label{directed::lem::correctness}
$Query(\lambda,u) \leq\delta_H(\lambda, u)+\eps\alpha$.
\end{restatable}
\defproof{dircorrect}{
\begin{proof}
Let $r$ be the root-most node in $\mathcal{T}_H$ such that the shortest $\lambda$-to-$u$ path $P$ in $H$ intersects some $Q\in Sep_r$. Let $k$ be a vertex in $Q\cap P$. By the definition of the connection set $C^{*}_{H_r}(S_{r}^{\lambda},Q)$, there exists $(q,\ell)$ such that
\begin{equation}
l+\delta_{H_r}(q,k)\leq\delta_{H_r}(S_{r}^{\lambda},k)+\eps^{*}\alpha
\end{equation}
 Also there exists $t\in C_{H_r}(Q,u)$ such that
\begin{equation}
\delta_{H_r}(k,t)+\delta_{H_r}(t,u)\leq\delta_{H_r}(k,u)+\hat\epsilon\alpha\leq\delta_{H_r}(k,u)+\eps^{*}\alpha
\end{equation}
We add the two to get
\begin{align*}
l+\delta_{H_r}(q,k)\delta_H(k,t)+\delta_{H_r}(t,u)&\leq\delta_{H_r}(S_{r}^{\lambda},k)+\delta_{H_r}(k,u)+\eps^{*}\alpha+\eps^{*}\alpha\\
l+\delta_{H_r}(q,t)+\delta_H(t,u)&\leq\delta_{H_r}(S_{r}^{\lambda},k)+\delta_{H_r}(k,u)+2\eps^{*}\alpha\\
l+\delta_{H_r}(q,t)+\delta_{H_r}(t,u) & \leq\delta_{H_r}(S_{r}^{\lambda},u)+\eps\alpha
\end{align*}
Clearly, $l+\delta_H(q,t)+\delta_H(t,u) \geq\delta_H(S_{r}^{\lambda},u)$. And since $P$ is fully contained in $H_r$, $\delta_{H_r}(S_{r}^{\lambda},u)=\delta_H(S_{r}^{\lambda},u)$, and the Lemma follows.
\qendproof}

\mysubsection{Update}
Assume that some vertex $u$ changes its label from $\lambda_1$ to $\lambda_2$. For every ancestor $r$ of $r_u$ and every $Q\in Sep_r$, we would like to remove $C_{H_r}(u, Q)$ from $C_{H_r}(S_{r}^{\lambda_1}, Q)$, and combine $C_{H_r}(u, Q)$ into $C_{H_r}(S_{r}^{\lambda_2},Q)$.
While the latter is straightforward using Lemma \ref{directed::lem::thinning_lemma}, removing $C_{H_r}(u, Q)$ from $C_{H_r}(S_{r}^{\lambda_1}, Q)$ is more difficult. For example, if $u$ was the closest $\lambda_1$  labeled vertex to every vertex on $Q$, it is possible that $C_{H_r}(u, Q)=C_{H_r}(S_{r}^{\lambda_1}, Q)$.
In that case, we will have to rebuild $C_{H_r}(S_{r}^{\lambda_1}, Q)$ from the other $O(|V(H_r)|)$ vertices of $S_{r}^{\lambda_1}$. Instead of removing the connections of $u$, we will rebuild $C_{H_r}(S_{r}^{\lambda_1}, Q)$ bottom-up starting from the leaf node $r_u$.

We therefore start by describing how to update $r_u$. There is a constant number of vertices in $r_u$, and hence $|S_{r_u}^{\lambda_1}|=O(1)$. Let $v_1$, $v_2$,...$v_k$ be the vertices in $S_{r_u}^{\lambda_1}$, such that $k=|S_{r_u}^{\lambda_1}|$. We stress that for every $1\leq j\leq k$, $v_j$ has an $\hat\epsilon$-covering set $C_{H_t}(v_j, Q)$ of size $O(\hat\epsilon^{-1})$ from $v_j$ to $Q$, for {\em every ancestor} $t$ of $r_u$ in $\mathcal T_H$, and for every $Q\in Sep_t$. 
We apply the Thinning Lemma (Lemma  \ref{directed::lem::thinning_lemma}) for each such $t$ and $Q$ on $\{C_{H_t}(v_j, Q)\}_{j=1}^k$ with $\eps'=\hat\epsilon$ and $\eps$ set to $\hat\epsilon$. Lemma \ref{directed::lem::thinning_lemma} yields a $2\hat\epsilon$-covering set $C_{H_t}(S^{\lambda}_{r_u}, Q)$.

We next handle the ancestors $r$ of $r_u$ in $\mathcal{T}_H$ in bottom up order. Let $x$ and $y$ be the children of $r\in\mathcal{T}_H$. We first note that $H_r$ = $H_x \cup H_y$ and hence, $S_{r}^{\lambda_1} = S_{x}^{\lambda_1} \cup S_{y}^{\lambda_1}$. 
Therefore, by 
Lemma \ref{directed::lem::thinning_lemma}, for every ancestor $t$ of $r$, and every $Q\in Sep_t$, $C_{H_t}(S_{r}^{\lambda_1},Q)$ can be obtained from $C_{H_t}(S_{x}^{\lambda_1},Q)\cup C_{H_t}(S_{y}^{\lambda_1},Q)$.
Let $i$ by the level of $r$ in $\mathcal{T}_H$, and hence the level of $x$ and $y$ is $i+1$. Since $t$ is an ancestor of $r$, it is also an ancestor of $x$ and $y$. Hence, $x$ ($y$) stores an $\eps_{i+1}$-covering set $C_{H_t}(S_{x}^{\lambda_1},Q)$ ($C_{H_t}(S_{y}^{\lambda_1},Q)$). We apply Lemma \ref{directed::lem::thinning_lemma} on $C_{H_t}(S_{x}^{\lambda_1},Q)$ and $C_{H_t}(S_{y}^{\lambda_1},Q)$ with $\eps'=\eps_{i+1}$ and $\eps=2\hat\epsilon$ to get an $(\eps_{i+1}+2\hat\epsilon)$-covering set $C_{H_t}(S_{r}^{\lambda_1},Q)$. The following lemma shows that $C_{H_t}(S_{r}^{\lambda_1},Q)$ is an $\eps_i$-covering set.

\begin{restatable}{lemma}{epsicover}
\label{lem:epsicover}
Let $r$ be a node in level $i$ in $\mathcal{T}_H$. For every ancestor $t$ of $r$, and every $Q\in Sep_t$, $C_{H_t}(S_{r}^{\lambda_1},Q)$ is an $\eps_i$-covering set from $S_{r}^{\lambda_1}$ to $Q$.
\end{restatable}

\defproof{epsicover}{
\begin{proof}
We first recall that $\eps_i = \frac{\eps\log{n}- i+1}{4\log{n}}$ for every $1\leq i \leq\log{n}$.
We prove the lemma by induction on the level of $r$ in $\mathcal{T}_H$. The base case is $i=\log{n}$, so $r$ is a leaf. The connection sets of the leaf nodes are computed explicitly using Lemma \ref{directed::lem::thinning_lemma}, with $\eps$ and $\eps'$ set to $\hat\epsilon$. Hence the product of the lemma is $2\hat\epsilon$-covering sets. 
\begin{align*}
2\hat\epsilon=2\frac{\eps}{8\log{n}}=\frac{\eps}{4\log{n}}=\eps_{\log {n}}
\end{align*}

For the inductive step, if $r$ is a leaf, then the arguments from the base case applies. Otherwise, let $x$ and $y$ be the children of $r$. By the induction hypothesis, both $x$ and $y$ have $\eps_{i+1}$-covering set from $S_{x}^{\lambda_1}$ and $S_{y}^{\lambda_1}$ to $Q$, respectively.
The update procedure applies Lemma \ref{directed::lem::thinning_lemma} on $C_{H_x}(S_{x}^{\lambda_1},Q)$ and $C_{H_y}(S_{y}^{\lambda_1},Q)$ with $\eps'=\eps_{i+1}$ and $\eps=2\hat\epsilon$, so we get an $(\eps_{i+1}+2\hat\epsilon)$-covering set $C_{H_t}(S_{r}^{\lambda_1},Q)$.
\begin{align*}
\eps_{i+1}+2\hat\epsilon=\eps\frac{\log{n}- (i+1)+1}{4\log{n}}+\eps\frac{2}{8\log{n}}=\eps\frac{\log{n}- i+1}{4\log{n}}=\eps_i
\end{align*}
\qendproof
}

To finish the update process, we need to update the $\eps^{*}$-covering sets that we use for queries. 
Let $r$ be an ancestor node of $r_u$ in level $i$ on $\mathcal{T}_H$. By Lemma~\ref{lem:epsicover}, for every $Q\in Sep_r$, we have an $\eps_i$-covering set $C_{H_r}(S_{r}^{\lambda_1},Q)$. Since $\eps_i < \eps^*$ , $C_{H_r}(S_{r}^{\lambda_1},Q)$ is also an $\eps^*$-covering set. However, it is too large. We apply Lemma \ref{directed::lem::thinning_lemma} on $C_{H_r}(S_{r}^{\lambda_1},Q)$ with $\eps'=\eps_i$, and $\eps$ set to $\frac{\eps}{4}$ to get $(\eps_i + \frac{\eps}{4})$-covering set. We note that since $\eps_i \leq\frac{\eps}{4}$ for every $1\leq i \leq \log{n}$, we get that $\eps_i +\frac{\eps}{4}\leq 2\frac{\eps}{4}\leq\frac{\eps}{2}=\eps^{*}$.
Hence the output of Lemma~\ref{directed::lem::thinning_lemma} is the desired $\eps^*$-covering set $C^{*}_{H_r}(S_{r}^{\lambda_1},Q)$.
We repeat the entire process for $\lambda_2$.

\begin{restatable}{lemma}{dirutime}
\label{directed::lem::update_time} 
There exists a scale-$(\alpha, \eps)$ distance oracle for directed $\alpha$-layered planar graph, with query time $O(\epsm\log{n})$ worst case, and update time of $O(\epsm \log^3{n})$ expected amortized. The oracle can be constructed in $O(\epsmm n\log^5{n})$ time and stored using $O(\epsm n\log^3{n})$ space.
\end{restatable}

\defproof{dirutime}{
\begin{proof}
Since our update process only uses Lemma \ref{directed::lem::thinning_lemma}, we bound the update time by the running time of that Lemma. Since the running time of Lemma \ref{directed::lem::thinning_lemma} is linear in sizes of the input connection sets we get the bound by the number of connection stored for $r_u$ and its ancestors. We store for $r_u$ $\frac{\eps}{4\log{n}}$-connection set for constant number of separators for every one of the $O(\log{n})$ ancestors of $r_u$. Hence, the number of connections stores for $r_u$ is $O(\log{n}(\frac{\eps}{4\log{n}})^{-1})=O(\epsm \log^2{n})$.
Since the number of connections stored for $r_u$ dominates the number of connection stored for any other strict ancestor of $r_u$, we get the total number of connections stored of $O(\log^3{n})$.

We note that the connections of the vertices in $r_u$ are only used when updating $r_u$, and for any other non-leaf node $r$, we only use the connection of its children. 
Thus, any connection is used at most twice. Once for updating a connection set of its parent, and the second time, is when updating the $\eps^{*}$-covering sets of $r$ ($r_u$). Hence the total input size of Lemma \ref{directed::lem::thinning_lemma} is at most twice the number of the connections stored for the ancestors of $r_u$, that is $O(\log^3{n})$, and the update time follows.

Since we store for every $r\in\mathcal T_H$ and every $Q\in Sep_r$ a connection set for every $\lambda\in \mathcal L_r$, we use dynamic hashing as in Section~\ref{sec::fast_query}. Hence, our update time is expected amortized.

Our query time is trivial and follows from the fact that we process $O(\log{n})$ levels in $\mathcal{T}_H$, and in each we inspect $O(\eps^{*-1})$ connections. That is $O(\epsm \log{n})$ time worst case.

By Lemma \ref{thorup::construction_time} with $\eps$ set to $\hat\epsilon$, all connection sets for all leaves of $\mathcal{T}_H$ can be computed in $O(\epsmm n\log^5{n})$ and it requires $O(\epsm n\log^2{n})$ space.
We construct the connection sets $C_{H_r}(S_{r}^{\lambda}, Q)$ for all $r\in\mathcal{T}_H$, $Q\in Sep_r$ and $\lambda\in H_r$ by applying the update process for each vertex $v\in V(H)$. This takes $O(\epsm\log^3{n})$ expected amortized time per operation, and $O(\epsm n\log^3{n})$ expected amortized time in total. This is dominated by the construction of Thorup's oracle.

To get the space requirements of our data structure, we need to count the number of connections stored. If every vertex $u\in V(H)$ has unique label, it follows that the connection sets stored for $u$ are not useful for any other vertex. We therefore count the number of $u$'s connections and multiple by $O(n)$.
Let $\lambda$ be the label of $u$. To support queries and updates for $\lambda$, we store for every ancestor $r$ of $r_u$ connection sets from $S_{r}^{\lambda}$ to $O(\log{n})$  separators for the ancestors of $r$. Since the size of these connection sets is only bounded by $O(\hat\eps^{-1})$, we get that $r$ requires $O(\epsm\log^2{n})$ connections. Since $r_u$ has $O(\log{n})$ ancestors, we store for $u$ (and by that for $\lambda$) $O(\log^3{n})$ connections. Thus, the total space required is $O(n\log^3{n})$. 
\qendproof
}

\noindent We can now apply Lemma \ref{thorup::lem::stretch_eps} to get the following theorem:
\begin{theorem}\label{thm::directed_structure}
For any directed planar graph and fixed parameter $\eps$, there exists a $(1+\eps)$ approximate vertex-labeled distance oracle that support queries in $O(\epsm\log{n}\lglg{nN})$ worst case and updates in $O(\epsm\log^3{n}\log{nN})$ expected amortized time. This oracle can be constructed in $O(\epsmm n\log^5{n}\log{nN})$ expected amortized time, and stored using $O(\epsm n\log^3{n}\log{nN})$ space.
\end{theorem}

\mysection[Undirected Graphs With Faster Update]{Oracle for Undirected Graphs With Faster Update} \label{sec::fast_update}
Both Thorup~\cite[Lemma 3.19]{Thorup04} and Klein~\cite{Klein02} independently presented efficient vertex-vertex distance oracles for undirected planar graph that use connections sets. Klein later improved the construction time~\cite{MSSP}.
They show that, in undirected planar graph, one can avoid the scaling approach using $\alpha$-layered graphs. Instead, there exist connections sets that approximate distance with $(1+\eps)$ multiplicative factor rather than $\eps\alpha$ additive factor.
We borrow the term \emph{portals} from Klein to distinct this type of connections from the previous type.
\begin{definition}
Let $G$ be an undirected planar graph, and let $Q$ be a shortest path in $G$. For every vertex $v\in V(G)$ we say that a set $C_G(v,Q)$ is an $\eps$-covering set of {\em portals} if and only if, for every vertex $t$ on $Q$ there exist a vertex $q$ on $Q$ such that: $\delta_G(v,q)+\delta_G(q,t)\leq(1+\eps)\delta_G(v,t)$
\end{definition}

We use a recursive decomposition $\mathcal T_G$ with shortest path separators, and use Klein's algorithm~\cite{MSSP} to select all the 
portal sets $C_{G_r}(u,Q)$ efficiently{\ifdefined \fullver \else{ (see appendix \ref{rmqconst_proof} for the exact time bounds)}\fi}. We cannot use the lists of Section~\ref{sec::fast_query} because there may be too many portals, and we cannot use the thinning
lemma (Lemma~\ref{directed::lem::thinning_lemma}) of Section~\ref{sec::directed} because its proof uses a directed construction, and hence, cannot be applied in undirected graphs. 
Instead, we take the approach used by Li, Ma and Ning for the static vertex-labeled case~\cite{LMN13}. We work with all portals of vertices with the appropriate label, and find the closest one using dynamic Prefix/Suffix Minimum Queries.

\begin{definition}[Dynamic Prefix Minimum Data Structure]\label{pmq} A Dynamic Prefix Minimum Data Structure is a data structure that maintains a set $A$ of $n$ pairs in $[1, n]\times \mathbb{R}$, under insertions, deletions, and Prefix Minimum Queries (PMQ) of the following form: given $l \in[1,n]$ return a pair $(x,y)\in A$ s.t. $x \in [1,l]$, and for every other pair $(x',y')$ with $x' \in[1,l]$, $y\leq y'$.
\end{definition}
Suffix minimum queries (SMQ) are defined analogously. Let $PMQ(A,l)$ and $SMQ(A,l)$ denote the result of the corresponding queries on set $A$ and $l$. 

We assume that for every $u,v\in V(G_r)$, $C_{G_r}(u,Q)\cap C_{G_r}(v,Q)=\emptyset$. This is without loss of generality, since if $x$ is a portal of a set of vertices ${v_0,...,v_k}$, we can split $x$ to $k$ copies. This does not increase |G| by more than a factor of $\epsm$ \ifdefined \fullver\ifdefined \isarticle \else(see figure \ref{fig:path_reduction})\fi.
\ifdefined \fullver
\begin{figure}
\begin{center}
\includegraphics[width=0.7\textwidth]{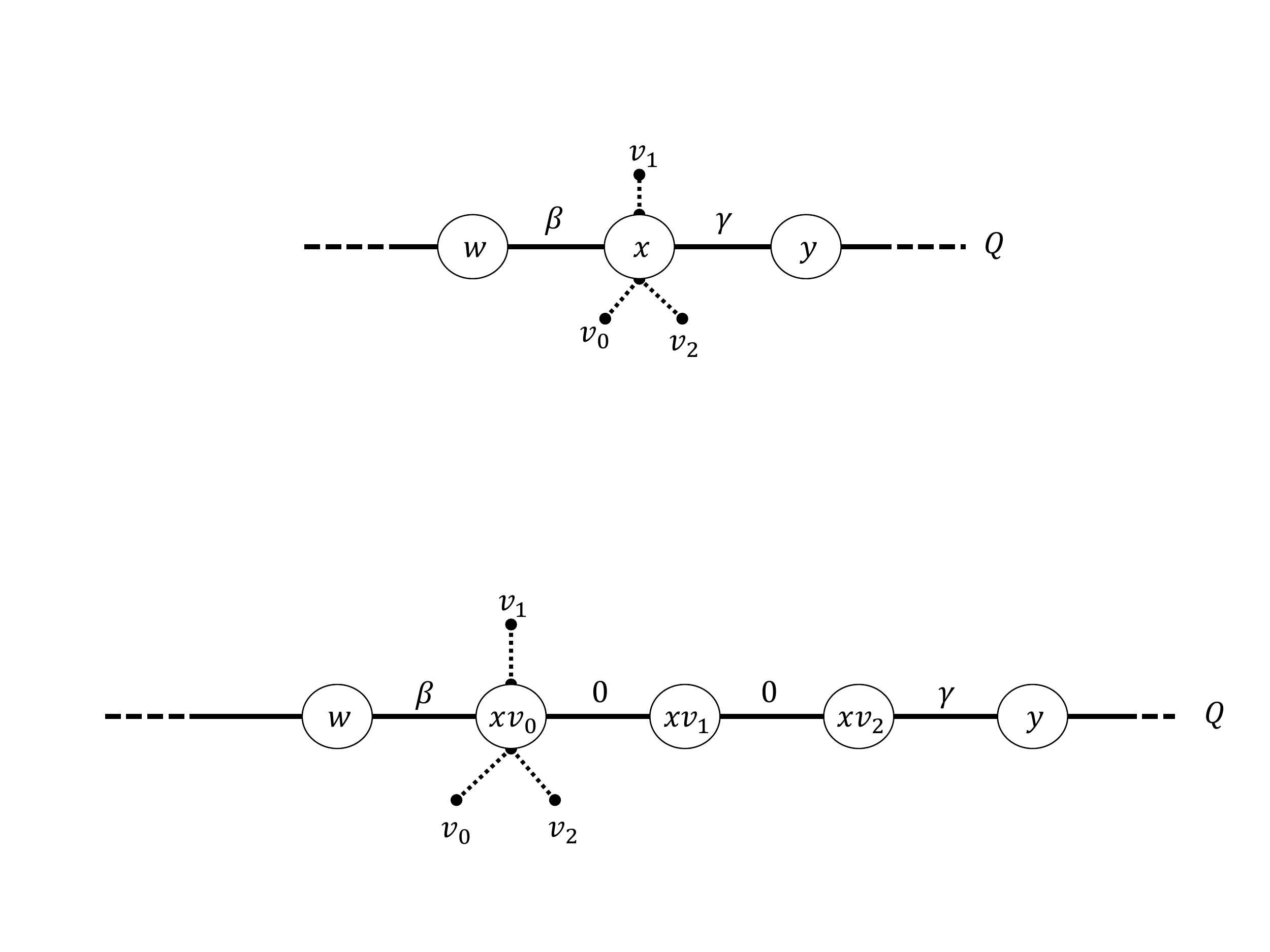}
\end{center}
\label{fig:path_reduction}
\caption{
Illustration of the reduction to unique portals. Above, the path $Q$ with the portal $x$ that is used by $v_0$, $v_1$, and $v_2$. Below, $x$ was replaced by $xv_0$, $xv_1$, and $xv_2$, inner connected with zero length edges. Here, $xv_0$, $xv_1$, $xv_2$ are the portals of $v_0$, $v_1$, and $v_2$ respectively. Note that this reduction does not introduce new paths in the graph, nor changes the distance along $Q$.
}
\end{figure}
\fi
To describe our data structure, we first need the following definitions. Let $Q\in Sep_r$ for some $r\in \mathcal{T}_G$.
Let $q_0, ..., q_k$ be the vertices on $Q$ by their order along $Q$. $G$ is undirected, hence the direction of $Q$ is chosen arbitrarily. For every $0\leq{j}\leq{k}$, let $h(q_j)$ denote the distance from $q_0$ to $q_j$ on $Q$. We note that since $Q$ is a shortest path in $G$, $h(q_i)=\delta_{G}(q_0,q_j)$. For every $\lambda\in\mathcal{L}_r$ we maintain a dynamic prefix minimum data structure $Pre_{Q,\lambda}$ over $\{ (j,-h(q_j)+\delta_{G_r}(q_j, \lambda))\}_{j=0}^k$. We similarly maintain a a dynamic sufix minimum data structure $Suf_{Q,\lambda}$ over $\{(j,h(q_j)+\delta_{G_r}(q_j, \lambda))\}_{j=0}^k$. 

\mysubsubsection{$Query(u,\lambda)$}
For every ancestor $r$ of $r_u$ in $\mathcal{T}_G$, every $Q\in Sep_r$, and every $q_j\in{C_{G_r}(u,Q)}$ we wish to find the index $i$ that minimizes $\delta_{G_r}(u,q_j)+\delta_{G_r}(q_j, q_i)+\delta_{G_r}(q_i, \lambda)$.
Observe that for $i\leq{j}$, $\delta_{G_r}(q_j, q_i)=h(j)-h(i)$, while for $i\geq{j}$, $\delta_{G_r}(q_j, q_i)=h(i)-h(j)$.
We therefore find the optimal $i\leq{j}$ and $i\geq j$ separately. Note that $min_{i\leq{j}}(\delta_{G_r}(u,q_j)+\delta_{G_r}
(q_j,q_i)+\delta_{G_r}(q_i, \lambda))=
\delta_{G_r}(u,q_j)+h(j)+PMQ(Pre_{Q,\lambda},j)$.
Similarly, we handle the case where $i\geq{j}$ using $SMQ(Suf_{Q,\lambda},j)$. Thus, we have two queries for each portal of $u$.
We also compute the distance from $u$ to $\lambda$ in $r_u$ explicitly. We return the minimum distance computed.

\begin{restatable}{lemma}{rmqcor}
\label{rmq::lem::correctness}
The query algorithm returns a distance $d$ such that 
$\delta_G(u,\lambda)\leq d \leq{(1+\eps)\delta_G(u,\lambda)}$
\end{restatable}

\defproof{rmqcor}{
\begin{proof}
The proof of correctness of our algorithm is essentially the same as in~\cite[Lemma 1]{LMN13}. We adapt it to fit our construction.
Let $v$ be the closest $\lambda$-labeled vertex to $u$ in $G$. If the shortest $u$-to-$v$ path $P$ does not leave $r_u=r_v$ the algorithm is correct, since the distance in $r_u$ is computed expilicitly. Otherwise, let $r$ be the root-most node in $\mathcal{T}_G$ such that $P$ intersects some $Q\in Sep_r$. 
Let $t$ be a vertex on $P\cap Q$. There exists $q_j\in C_{G_r}(u,Q)$ and $q_i\in C_{G_r}(v,Q)$ such that:
\begin{align}
\delta_{G_r}(u,q_j)+\delta_{G_r}(q_j,t)\leq (1+\eps)(\delta_{G_r}(u,t))\\
\delta_{G_r}(v,q_i)+\delta_{G_r}(q_i,t)\leq (1+\eps)(\delta_{G_r}(v,t))
\end{align}
We add the two inequalities to get the following:
\begin{align}
\delta_{G_r}(u,q_j)+\delta_{G_r}(q_j,q_i)+\delta_{G_r}(v,q_i)\leq (1+\eps)(\delta_{G_r}(u,v))\label{undirected::fast_update::correctness::eq_1}
\end{align}

If $i\leq{j}$, then $PMQ(Pref_{Q,\lambda},j)\leq\delta_{G_r}(q_i,\lambda)-h(q_i)\leq\delta_{G_r}(v,q_i)-h(q_i)$. Thus,
\begin{align}
Query(u,\lambda)&\leq\delta_{G_r}(v,q_i)-h(q_i) + h(j)+\delta_{G_r}(u,q_j)\\
&=\delta_{G_r}(v,q_i)+\delta_{G_r}(q_i,q_j)+\delta_{G_r}(u,q_j)\\
&\leq(1+\eps)(\delta_{G_r}(u,v))\label{undirected::fast_update::correctness::eq_2}\\
&\leq(1+\eps)(\delta_{G}(u,\lambda))\label{undirected::fast_update::correctness::eq_3}
\end{align}
Here, inequality (\ref{undirected::fast_update::correctness::eq_2}) follows from (\ref{undirected::fast_update::correctness::eq_1}), and (\ref{undirected::fast_update::correctness::eq_3}) follows from that fact that $P$ is fully contained in $r$, and our assumption that $v$ is the closest $\lambda$-labeled vertex to $u$.

The proof for the case that $i\geq{j}$ is similar.
\qendproof
}

\mysubsubsection{Update}
Assume that the label of $u$ changes from $\lambda_1$ to $\lambda_2$. For every ancestor $r$ of $r_u\in \mathcal{T}_G$, and $Q\in Sep_r$, and for $q_i\in C_{G_r}(u,q)$, we remove from $Pre_{Q,\lambda_1}$ and $Suf_{Q,\lambda_1}$ the element $(x,y)$ with $x=i$, and insert the element $(i,-h(i)+\delta_{G_r}(u,q_i))$ into $Pre_{Q,\lambda_2}$, and $(i,h(i)+\delta_{G_r}(u,q_i))$ into $Suf_{Q,\lambda_2}$.
We note that since we assume that every vertex $q_i$ is a portal of at most one vertex, the removals are well defined, and the insertions are safe.

The time and space bounds for the oracle described above are given in the following lemma.
\begin{restatable}{lemma}{rmqconst}
\label{rmq::lem::construction}
Assume there exists a dynamic prefix/suffix minimum data structure that, for a set of size $m$, supports PMQ/SMQ in $O(T_Q(m))$ time, and updates in  $O(T_U(m))$ time, can be constructed in $O(T_C(m))$ time, where $T_C(m)\geq m$, and can be stored in $O(S(m))$ space. Then there exist a dynamic vertex-labeled stretch-$(1+\eps)$ distance oracle for planar graphs with worst case query time
$O(\epsm \log(n) T_Q(\epsm n))$, 
and expected amortized update time $O(\epsm\log(n)T_U(\epsm n))$. The oracle can be constructed using  
$O(n\log^2n + \log(n)T_c(\epsm n))$ expected amortized time, 
and stored in $O(\log(n) S(\epsm n))$  space.
\end{restatable}

\defproof{rmqconst}{
\begin{proof}
\label{rmqconst_proof}
Let $G$ be an undirected planar graph. We first decompose $G$ to obtain $\mathcal T_G$, and compute all the portals and the distances to portals. Klein~\cite{MSSP} shows that this can be done using 
$O(n\log(n)(\epsm+\log n))$ time.
Then, for every $r\in\mathcal{T}_G$, for every $Q\in Sep_r$ and every $\lambda\in\mathcal{L}_r$, we construct a prefix/suffix minimum query data structures for $Pre_{Q,\lambda}$ and $Suf_{Q,\lambda}$. This takes $\log(n) T_C(\epsm n)$ time, since at every level of $\mathcal T_G$ the total number of portals is $\epsm n$, and since $T_C(\cdot)$ is superlinear.
The number of portals we store is $O(\epsm n\log n)$ since every vertex $v$ has $O(\epsm)$ portals for every one of its $O(\log n)$ ancestors in $\mathcal{T}_G$. Hence our space is $O(\log(n)S(\epsm n))$, and the construction time is $O(n\log^2n + \log(n)T_c(\epsm n))$.

To analyze the query and update time, we note that we process $O(\log n)$ nodes in $\mathcal{T}_G$ and in each we perform $O(\epsm)$ queries or updates to the prefix/suffix minimum query structures. The size of our prefix/suffix structures is bounded by the size of $V(Q)$ which is $O(\epsm n)$. The $\epsm$ factor is due to the assumption of distinct portals.
Thus, the query time is $O(\epsm\log(n)T_Q(\epsm n))$ and the update time is $O(\epsm\log(n)T_U(\epsm n))$.

Since every $Q\in Sep_r$ holds a prefix/suffix minimum data structure for every label $\lambda\in\mathcal{L}_r$, we use dynamic hashing to avoid space dependency in $|\mathcal{L}|$, as in Section \ref{sec::fast_update}. Hence, our construction time and update time are expected amortized.
\qendproof
}

It remains to describe a fast prefix/suffix minimum query structure.
\@namedef{AppendixB}{
We use a result due to Wilkinson~\cite{Wilkinson} for solving the 2-sided reporting problem. In this problem, we maintain a set $A$ of n points in $\mathbb{R}^2$ under an online sequence of insertions, deletions and queries of the following form. Given a rectangle $B=[l_1,h_1]\times[l_2,h_2]$ such that exactly one of ${l_1, l_2}$ and one of ${h_1, h_2}$ is $\infty$ or $-\infty$, we report $A\cap B$.
Here, $[l_1,h_1]\times[l_2,h_2]$ represents the rectangle $\{(x,y) : l_1\leq x \leq l_2 , h_1 \leq y \leq h_2 \}$. 
Wilkinson's data structure is captured by the following theorem.
\begin{theorem}\cite[Theorem 5]{Wilkinson} For any $f\in[2, \log{n}/\lglg{n}]$, there exists a data structure for 2-sided reporting with update time $O((f\log{n}\lglg{n})^{1/2})$, query time $O((f\log{n}\lglg{n})^{1/2}+\log_f(n)+k)$ where $k$ is the number of points reported. The structure requires linear space.
\end{theorem}
In fact, Wilkinson's structure first finds the point with the minimum $y$-coordinate in the query region, and then reports the other points. Using this fact, and setting $f=\log^{\gamma}{n}$ for some arbitrary small constant $\gamma$. We get the following lemma. We also state Wlikinson's construction time explicitly.
\begin{lemma}\label{undirected::fast_update::wilkinson::restate}
There exists a linear space data structure that maintains a set of $n$ points in $\mathbb{R}^2$, with update time $O(\log^{{1/2}+\gamma}{n})$, that given a 2-sided query, returns the minimum $y$-coordinate of a point in the query region in $O(\frac{log{n}}{\lglg{n}})$ time. This data structure can be constructed in $O(n\log^{{1/2}+\gamma}{n})$.
\end{lemma}
Our prefix/suffix queries correspond to one-sided range reporting in the plane, which can be solved using 2-sided queries, by setting the upper limit of the query rectangle to $nN$.
\begin{lemma}
\label{undirected::fast_update::existence_of_psq}
For any constant $\gamma > 0$, there exists a linear space dynamic prefix/suffix minimum data structure over $n$ elements with update time $O(\log^{{1/2}+\gamma} n)$, and query time $O(\frac{log{n}}{\lglg{n}})$. This data structure can be constructed in $O(n\log^{{1/2}+\gamma} n)$ time.
\end{lemma}

\begin{proof}
We regard each element $(i,l)$ of $A$ as a point in $\mathbb R^2$, and use Wilkinson's structure. 
A prefix minimum query for $i$ corresponds to finding the point with minimum $y$-coordinate in the rectangle $(-\infty,-\infty,i,\infty)$.
This is 1-sided rectangle. To be able to specify a boundry for the $y$-axis, we maintain an upper bound $y_{max}$ on the $y$-coordinates of points in $A$. The bound can be easily updated in constant time when an insertion occurs. (There is no need to update the bound when a deletion occurs). 
We replace the 1-sided rectangle with the 2-sided rectangle $(-\infty,-\infty,i,y_{max})$. Similarly, our suffix minimum query is the 1-sided rectangle $(i,-\infty,\infty,\infty)$ or the 2-sided $(i,-\infty,\infty, y_{max})$. The lemma now follows by applying Lemma \ref{undirected::fast_update::wilkinson::restate}.
\qendproof
}

We use a result due to Wilkinson~\cite{Wilkinson} for solving the 2-sided reporting problem, from which a prefix/suffix minimum data structure easily follows. This is summarized in the following lemma. See Appendix~\ref{app:Wilkinson} for the full details.

\begin{lemma}
\label{undirected::fast_update::existence_of_psq}
For any constant $\gamma > 0$, there exists a linear space dynamic prefix/suffix minimum data structure over $n$ elements with update time $O(\log^{{1/2}+\gamma} n)$, and query time $O(\frac{log{n}}{\lglg{n}})$. This data structure can be constructed in $O(n\log^{{1/2}+\gamma} n)$ time.
\end{lemma}

\noindent We therefore obtain the following theorem.
\begin{theorem}\label{thm::fast_update}
For any undirected planar graph and fixed parameters $\eps, \gamma$, there exists a stretch-$(1+\eps)$ vertex-labeled distance oracle that approximates distances in $O(\epsm\frac{\log n\log(\epsm n)}{\lglg(\epsm n)})$ time worst case, and supports updates in
$O(\epsm\log n\log^{\frac{1}{2}+\gamma}(\epsm n))$ expected amortized time. This data structure can be constructed using 
$O(n\log^2 n +$\\ $\epsm n \log n \log^{\frac{1}{2}+\gamma}{(\epsm n)})$ 
expected amortized time and stored using $O(\epsm n\log n)$ space.

\end{theorem}
	\mysection{Conclusion}
	In this paper we presented approximate vertex-labeled distance oracles for directed and undirected planar graphs with polylogarithmic query and update times and nearly linear space.
	All of our oracles have  $\Omega(\log{n})$ query and updates since we handle root-to-leaf paths in the decomposition tree. It would be interesting to study whether this can be avoided, as done in the vertex-to-vertex case, where approximate distance oracles with faster query times exist (see e.g.,~\cite{Thorup04,WulffN16,GuX15} and references therein). 
	Another interesting question that arises is that of faster dynamic prefix minimum data structures. In Section~\ref{sec::fast_update} we used Wilkinson's 2-sided reporting~\cite{Wilkinson} as a dynamic prefix/suffix minimum data structure. Can other approaches to this problem be used to obtain a faster solution?
	
	\mysection*{Acknowledgements}
	We thank Pawe{\l} Gawrychowski and Oren Weimann for fruitful discussions.

\bibliography{dynamic}

\newpage
\appendix
\mysection*{Appendix}\label{apndx}

\mysection[Reduction from stretch-$(1+\eps)$ to scale-$(\alpha, \eps)$ distance oracle]{Reduction from stretch-$(1+\eps)$ vertex-labeled distance oracle to scale-$(\alpha, \eps)$ distance oracle.} \label{apndx::scale_to_stretch_reduction}
We now describe how to use a scale-$(\eps, \alpha)$ vertex-labeled distance oracle to obtain a stretch-$(1+\eps)$ distance oracle.
For vertex-vertex distance oracles, this reduction was proven by Thorup as captured in Lemma \ref{thorup::lem::stretch_eps}.

For the static vertex-labeled case, a similar reduction was presented by Mozes and Skop, and is as follows:
The proof of Lemma \ref{thorup::lem::stretch_eps} relies on two reductions~\cite[Lemmas 3.2,3.8]{Thorup04}. The first shows that from any graph $G$ and for any $\alpha > 0$, one can construct a family of $\alpha$-layered graphs $\{G^\alpha_i\}_i$ whose total size is linear in the size of $G$, and such that:

\begin{enumerate}
\begin{item}\label{apndx::scale_to_stretch_reduction::item_1}
$\Sigma |G_i^\alpha|=O(|G|)$, where $|G|=|V(G)|+|E(G)|$.
\end{item}
\begin{item}\label{apndx::scale_to_stretch_reduction::item_2}
Each $v\in V(G)$ has an index $j(v)$ s.t. any $w\in V(G)$ has $d=\delta_G(v,w)\leq\alpha$ iff $d=min\{\delta_{G^\alpha_{j(v)-2}}(v,w), \delta_{G^\alpha_{j(v)-1}}(v,w), \delta_{G^\alpha_{j(v)}}(v,w)\}$.
\end{item}
\begin{item}\label{apndx::scale_to_stretch_reduction::item_3}
Each $G^\alpha_i$ is a minor of $G$. I.e., it can be obtained from $G$ by contraction ad deletion of arcs and vertices. In particular, if $G$ is planar, so is $G^\alpha_i$.
\end{item}
\end{enumerate}
Item (\ref{apndx::scale_to_stretch_reduction::item_2}.) means that any shortest path of length at most $\alpha$ in $G$ is represented
in at least one of three fixed graphs $G_i^\alpha$. Thus, one can use scale-$(\alpha, \eps)$ distance oracles for the $\alpha$-layered graphs $\{G_i^\alpha\}$ to implement a scale-$(\alpha, \eps)$ oracle of $G$.

The second reduction~\cite[Lemmas 3.8]{Thorup04} is a scaling argument that shows how to construct a stretch-$(1+\eps)$ distance oracle for $G$ using scale-$(\alpha, \eps')$ distance oracles for $\alpha\in\{2^i\}_{i\in[1, \lceil \log(nN \rceil]}$. The reduction does not rely on planarity.
Now consider the vertex-labeled case. Let $G^*$ be the graph obtained from $G$ by adding apices representing the labels. A vertex-to-vertex distance oracle for $G^*$ is a vertex-labeled distance oracle for $G$, and vice versa. By Thorup's second reduction, it suffices to show how to construct a scale-$(\alpha, \eps)$ vertex-vertex distance oracle for $G^*$ for any $\alpha$, $\eps$, or, equivalently a  vertex-labeled scale-$(\alpha, \eps)$ distance oracle for $G$ and every $\alpha$, $\eps$. Let $\alpha\in\mathbb{R}^+$. Given $u\in V(G)$ and $\lambda\in\mathcal{L}$ with $\delta_G(u,\lambda)\leq\alpha$, let $w\in V(G)$ be the closest $\lambda$ labeled vertex to $u$. By the properties of Thorup's first reduction, there is a graph $G_i^\alpha$ in whitch the $u$-to-$w$ distance is $\delta_G(u,w)$. Thus, a vertex-labeled distance oracle for $G_i^\alpha$ will report a distance of at most $\delta_G(u,\lambda)+\eps\alpha$. Therefore we have the following Lemma:
\begin{lemma}\label{apndx::scale_to_stretch_reduction::query_lemma}
For any planar graph $G$ and fixed parameter $\eps$, a stretch-$(1+\eps)$ vertex-labeled distance oracle can be constructed using $O(\log{nN})$ scale-$(\alpha,\eps')$ vertex-labeled distance oracles where $\alpha=2^i$, $i=0,...\lceil{\log{nN}}\rceil$ and $\eps'\in{1/2, \eps/4}$.
Assume that the scale-$(\alpha, \eps)$ vertex-labeled distance oracle supports queries in $O(T_Q(n, \eps))$ and updates in $O(T_U(n, \eps))$ time, and it can be constructed in $O(T_C(n,\eps))$ time and uses $O(S(n,\eps))$ space. There exists a $S(n,\eps)\log{nN}$ space stretch-$(1+ \eps)$ vertex-labeled distance oracle that answers queries in $O(T_Q(n, \eps)\lglg(nN))$ and updated in $O(T_U(n, \eps)\log{nN})$ time can be constructed in $O(T_C(n,\eps)\log{nN})$ time.
\end{lemma}
\begin{proof}
Given a planar graph $G$, we decompose $G$ to $O(\log{nN)}$ $\alpha$-layered graphs, and for each we construct a scale-$(\alpha, \eps)$ distance oracle. We get the space requirements, and the construction and query times by using Lemma \ref{thorup::lem::stretch_eps}. Since we must keep all $O(\log{nN})$ scale oracles up to date, we perform each update operation $O(\log{nN})$ times, and the lemma follows.
\qendproof

\ifdefined \fullver 
\else {
	\mysection{Omitted Proofs}\label{sec::omitted_proofs}
	\mysubsection{Dynamic Vertex Labeled Distance Oracle for Undirected Graphs With Fast Query Time}
		\showproof{constconnectionset}
		\showproof{fastquerycorrect}
		\showproof{constutime}
		\showproof{constspace}

	\mysubsection{Dynamic Vertex Labeled Distance Oracle for Directed Graphs}
		\showproof{setsize}
		\showproof{thinninglem}
		\showproof{dircorrect}
		\showproof{epsicover}
		\showproof{dirutime}

	\mysubsection{Dynamic Vertex Labeled Distance Oracle for Undirected Graphs With Fast Update}
		\showproof{rmqcor}
		\showproof{rmqconst}
}
\fi

\mysection{From 2-sided range reporting to prefix minimum queries}\label{app:Wilkinson}
\@nameuse{AppendixB}

\end{document}